\documentclass[doublespacing]{elsart}

\usepackage{icarus}

\usepackage{natbib}
\bibpunct{(}{)}{;}{a}{,}{,}
 \usepackage{epsfig}
 \usepackage{amssymb}

\newcommand{\BibTeX}{ \textrm{B\kern-.05em\textsc{i\kern-.025em b}\kern-.08em
    T\kern-.1667em\lower.7ex\hbox{E}\kern-.125emX} }

\begin{document}

\begin{frontmatter}

\title{Population control of Mars Trojans by the Yarkovsky \& YORP effects}

\author[addr1]{Apostolos A.~Christou}
\author[addr1,addr2]{Galin Borisov}
\author[addr3]{Aldo Dell'Oro}
\author[addr4]{Seth A.~Jacobson}
\author[addr5]{Alberto Cellino}, and
\author[addr6]{Eduardo Unda-Sanzana}

\address[addr1]{Armagh Observatory and Planetarium, College Hill,
           Armagh BT61 9DG, Northern Ireland, UK}
\address[addr2]{Institute of Astronomy and National Astronomical Observatory, Bulgarian Academy of Sciences, 72, Tsarigradsko Chauss\'{e}e Blvd., BG-1784 Sofia, Bulgaria}
\address[addr3]{INAF - Osservatorio Astrofisico di Arcetri,
			 Largo Enrico Fermi 5, 
			 I-50125 Florence,
			Italy}
\address[addr4]{Department of Earth and Planetary Sciences, Northwestern University, Evanston, IL 60208-3130, USA}			
\address[addr5]{INAF - Osservatorio Astronomico di Torino,
			Via Osservatorio 20,
			Pino Torinese,
			10025 Torino,
			Italy} 		          
\address[addr6]{Centro de Astronom\'{i}a (CITEVA), Universidad de Antofagasta, Avenida Angamos 601 Antofagasta, Chile}

\end{frontmatter}

\begin{flushleft}
\vspace{1cm}
Number of pages: \pageref{lastpage} \\
Number of tables: \ref{tab:sign_0p56}\\
Number of figures: \ref{fig:joint_pdf_k1}\\
\end{flushleft}

\begin{pagetwo}{Population control of Mars Trojans by YORP and Yarkovsky}

Apostolos Christou \\
Armagh Observatory and Planetarium\\
College Hill\\
BT61 9DG Armagh\\
Northern Ireland, UK\\
\\
Email: apostolos.christou@armagh.ac.uk\\
Phone: +44 (0)28 3752 2928 \\
Fax: +44 (0)28 3752 7174

\end{pagetwo}

\begin{abstract}
We explore the hypothesis that the population of Martian Trojans is the result of a balance between the production of new asteroids (``YORPlets'') through the YORP effect and their eventual escape from the Trojan clouds through Yarkovsky-driven orbital evolution. Our principal observables are: (5261) Eureka, its family of 8 asteroids and the family-less Trojans (101429) 1998 $\mbox{VF}_{31}$ \& (121514) 1999 $\mbox{UJ}_{7}$. We model the population evolution as a birth-death random process and assume it is in a steady state. We then simulate the discovery of Trojans to-date and find that family members of 101429 and 121514, if they exist, are intrinsically more difficult to detect than Eureka's. Their non-discovery can be used as evidence of their non-existence only under the assumption that their brightness relative to the parent asteroid is similar to that in the Eureka family. To find out how efficiently Mars Trojans are lost from the Trojan clouds due to the Yarkovsky effect, we carry out dynamical simulations of test particles originating from these parent bodies. We find that objects originating from Eureka and 121514 begin escaping after $\sim$1 Gyr, but that those from 101429 are already lost by that time, probably due to that asteroid's proximity to an eccentricity-type secular resonance within Mars's co-orbital region. This is the likely cause behind the absence of Trojans in the orbital vicinity of 101429. In contrast, the solitary status of 121514 points to an intrinsic inability of the asteroid to produce YORPlets during the most recent $\sim$20\% of the solar system's history, a finding potentially related to 121514's present, low angular momentum rotational state, unless the Eureka family formed rapidly during a single fission event.
\end{abstract}


\section{\label{sec:intro}Introduction}
Trojan asteroids orbit the Sun some $60^{\circ}$ ahead or behind a planet's position along its orbit \citep{MurrayDermott1999}.
Mars is the only terrestrial planet known to host stable Trojans \citep{Scholl.et.al2005,Dvorak.et.al2012}. Recently, it was recognised that the orbits of several Mars Trojan asteroids form a family together with (5261) Eureka \citep{Christou2013,deLaFuenteMarcoses2013} indicating a genetic relationship between its members. Subsequent spectroscopic observations confirmed that the surfaces of family members have similar olivine-rich composition \citep{Borisov.et.al2017,Polishook.et.al2017} leading \citeauthor{Polishook.et.al2017} to suggest that they may be pieces of the Martian mantle, excavated from the planet's interior by a giant impact early in its history, and stored in its Trojan clouds.

Here we focus on another feature of the Martian Trojan population, namely that the three largest Trojans - (5261) Eureka \& (101429) 1998 $\mbox{VF}_{31}$ at $\mbox{L}_{5}$ and (121514) 1999 $\mbox{UJ}_{7}$ at $\mbox{L}_{4}$ - are each 1-2 km across, mineralogically distinct \citep{Rivkin.et.al2003,Rivkin.et.al2007} and in low-eccentricity orbits at 1.52 au, yet only Eureka is associated with a family. The typical timescale for catastrophic disruption due to impact of these objects is $>$ 6 Gyr and typically 10 Gyr for a 2 km Trojan \citep{Christou.et.al2017}. The formation of a family from the catastrophic disruption of one out of three over the age of the solar system is therefore a plausible, if still unlikely,  outcome if we consider the collisional history of these objects as independent repeats of the same natural experiment. On the other hand, the strong orbital compactness of the observed family is incompatible with a set of collisional fragments \citep{Christou.et.al2017}, and the spreading of the orbits due to Yarkovsky suggests an age of $\sim$1 Gyr \citep{Cuk.et.al2015}, which is difficult to reconcile with the long collisional lifetime. Invoking a low-energy collision (i.e.~a cratering event) overcomes the compactness problem, but it would create a family size distribution different from that observed \citep{Christou.et.al2017}. Therefore, either the family forming event was a statistical fluke or it formed in a different way. 

The Yarkovsky-O'Keefe-Radzievskii-Paddack (YORP) effect can produce new asteroids by spin-up and rotational fission of a parent body \citep{Walsh.et.al2008,JacobsonScheeres2011}, a mechanism also responsible for close orbital pairs \citep{Pravec.et.al2010} and clusters \citep{Pravec.et.al2018} of small asteroids in the Main Belt, and one that becomes more efficient closer to the Sun \citep{Jacobson2014}. Clues to the importance of YORP in this case come from the degree of similarity between the libration amplitudes of Eureka and its family members around $\mbox{L}_{5}$. These amplitudes are relatively immune to change by the Yarkovsky effect \citep{Cuk.et.al2015} and a good proxy for the original orbits of the family members. For the Eureka family, the spread in amplitudes translates into a velocity spread of $<$ 3 m $\mbox{s}^{-1}$ \citep{Cuk.et.al2015,Christou.et.al2017}, comparable to the escape velocity from Eureka ($\sim 1$ m $\mbox{s}^{-1}$ for a bulk density $\rho=1-3$ g $\mbox{cm}^{-3}$). This implies a gentle separation from the primary which is consistent with the YORP fission and escape scenario but more difficult to reconcile with an impact origin. The conclusion of a YORP-driven process is reasonable as pointed out by \citet{Christou2013}, since the estimated small mass ratios ($<$0.2) between Eureka and its family members are also observed in other asteroid clusters in the inner main belt, thought to arise from the rapid escape of the secondary into heliocentric orbit soon after fission \citep{Scheeres2009,Pravec.et.al2010}. Furthermore, Eureka's fast rotation rate \cite[$P$=2.69 hr;][]{Koehn.et.al2014} is right at the so-called ``spin barrier'' \citep{Warner.et.al2009}. 

This paper brings together currently available information on the orbital and physical properties of these three Trojans and examines whether, and under which conditions, a family associated with each of the Trojans {\it should} exist in a scenario where new Trojans are continuously created but also lost.  We aim to investigate why Eureka is the only Trojan with a family and estimate the likelihood of this.

The next two Sections of the paper set the scene for this investigation. We summarise our current state of knowledge of the Trojans, including newly-discovered members of the Eureka family, the latest information on their rotation periods and discuss the timescales relevant to Mars Trojan physical evolution. In Section~\ref{sec:random} we introduce a model of the Martian Trojans as the steady state in a birth-death random process and in Section~\ref{sec:escape} we model their loss from the Trojan clouds through the action of the Yarkovsky force. Section~\ref{sec:existence} investigates the statistical significance of the current absence of family members for the two other Trojans, used to define the observable in this study. In Section~\ref{sec:constraints} we translate our model results into constraints for the production rate of new Martian Trojans from the different parent asteroids. Section~\ref{sec:end} summarises our conclusions and suggests avenues for further investigation. 

\section{\label{sec:eureka}Current knowledge on the Martian Trojans}
Table~\ref{tab:troj} shows currently available information on the Trojans. For the purposes of this work, we distinguish
between our potential parent bodies - compositionally distinct asteroids that occupy different corners of the Trojan clouds - and the smaller asteroids belonging to the Eureka family. \citet{Christou.et.al2017} lists 6 family member asteroids apart from Eureka itself. To this we now add 2016 $\mbox{CP}_{31}$ and 2011 $\mbox{SP}_{189}$, first detected on previous apparitions and recovered in early 2018. We have carried out numerical simulations to confirm that they are stable $\mbox{L}_{5}$ Trojans of Mars and that their orbital evolution is similar to that of previously recognised family members. 2011 $\mbox{SP}_{189}$ is also the faintest member so far identified, with $H=20.9$. For the remainder of the paper we will ignore the offspring and refer to the individual parent asteroids as 5261 (or Eureka), 101429 and 121514.

Apart from mineralogical interpretation of reflectance spectra, available physical information of the Trojans is limited to bulk properties such as size, albedo and rotation period for the largest asteroids. The rotation period of 101429 was estimated to be $P=7.70$ hr \citep{Borisov.et.al2016} using Lomb-Scargle period analysis, while Eureka's is $\sim$2.7 hr. More recently, \citet{Borisov.et.al2018} found $P\simeq 46$ hr for 121514, considerably slower than either Eureka or 101429. 

\begin{figure}
\includegraphics[width=90mm,angle=0]{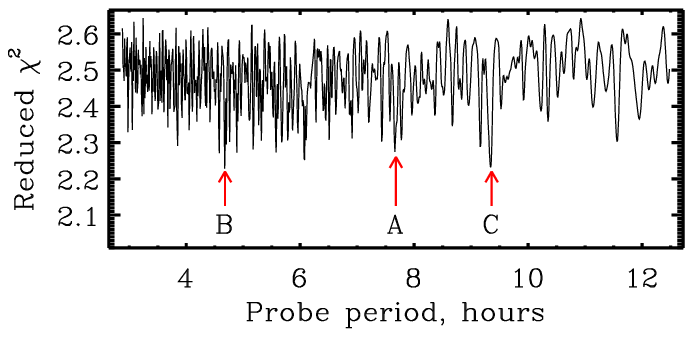}\\
\includegraphics[width=90mm,angle=0]{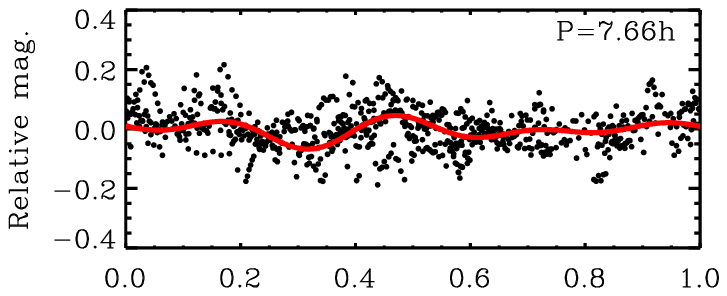}\\
\includegraphics[width=90mm,angle=0]{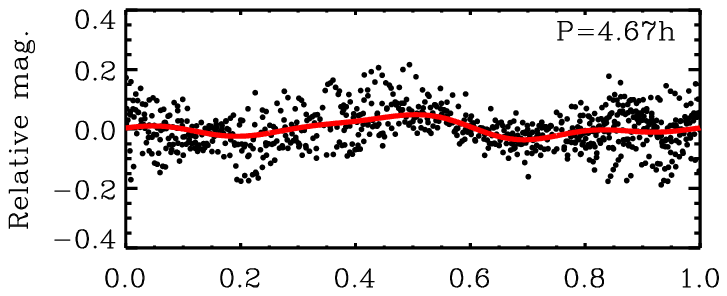}\\
\includegraphics[width=90mm,angle=0]{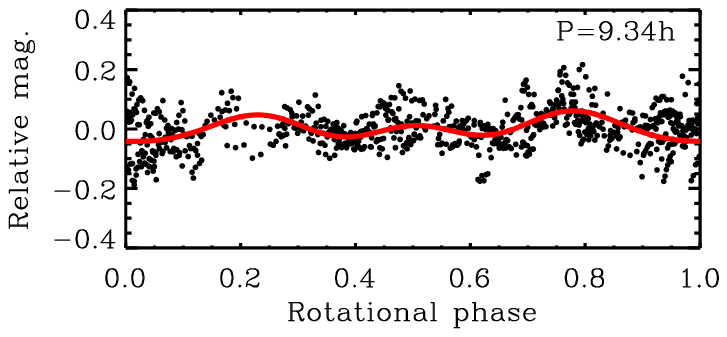}
\caption[Rotation period estimates and lightcurve fits for 101429. The top panel represents the variation of the $\chi^2$ goodness-of-fit for different probe periods (see text for details). Fourier fits for the three solutions marked with arrows are presented in the remaining panels. From top to bottom: ``A" \citep{Borisov.et.al2016}, ``B'' and ``C'' (respective best and second-best solution from the present analysis).]{Rotation period estimates and lightcurve fits for 101429. The top panel represents the variation of $\chi^2$ goodness-of-fit for different probe periods (see text for details). Fourier fits to the three solutions marked with arrows are presented in the remaining panels. From top to bottom: ``A" \citep{Borisov.et.al2016}, ``B'' and ``C'' (respective best and second-best solution from the present analysis).}
\label{fig:101429_lc}
\end{figure}

 We have re-processed the photometric data for 101429 using the method in the more recent work for 121514 where the data is fit to a Fourier function. This is done for consistency with \citet{Borisov.et.al2018} and because the new method allows more control over the fitting procedure than was available with the software tools used in \citet{Borisov.et.al2016}. We apply the procedure while varying the probe period with a step of 0.01\,hr and compute the reduced $\chi^2$ of the fit for each trial value. This is shown in Fig.~\ref{fig:101429_lc} for periods between 3 and 12.5\,hr. Our new best solution for 101429 is $P=4.67$ hr, corresponding to  the minimum marked ``B'' (top panel). A slightly worse solution, but still better than the previous solution ``A'', is $P=9.34$ hr (marked ``C''). It is worth pointing out the relatively low amplitude (0.2${}^{\rm m}$) of the lightcurve, which rules out a very elongated shape and a nearly equatorial view at the observation epoch. Unfortunately, even this new analysis does not succeed in constraining the period significantly better than in the work by \citet{Borisov.et.al2016}. New, higher quality data will be required to achieve this. 

\section{\label{sec:time}Timescales}
Processes that bear on the problem of Mars Trojan evolution are: (a) collisional spin reset (b) collisional disruption, and (c) YORP-induced disruption. Timescales for (b) and (c) were determined, the former specifically for Mars Trojans in \citet{Christou.et.al2017} and the latter in the general context of asteroid physical evolution in \citet{Jacobson.et.al2014}. In the following we examine the remaining process (a), that of collisional spin reset. As our modelling approach shares common elements with \citeauthor{Christou.et.al2017}, we provide a summary of our methodology below and refer the reader to Section~5 of that paper for the details.

As per \citet{Farinella.et.al1998}, a Òspin resetÓ is defined to occur when a projectile hitting the target imparts a rotational angular momentum equal or larger than the target angular momentum. In our modelling we do not consider the momentum-draining effect of impact ejecta preferentially escaping in the direction of rotation \citep{DobrovolskisBurns1984} because our aim is to estimate how often individual collisions reset an asteroid's angular momentum budget. For a spherical target of diameter $D_{\rm T}$, density $\rho_{\rm T}$ and rotation period $P$, the minimum diameter $D$
of a projectile of density $\rho$ and impact speed $U$ to produce a spin reset is
\begin{equation}
D= (3/10)^{1/3} (\rho_{\rm T}/\rho)^{1/3} (\omega D_{\rm T} /U)^{1/3} D_{\rm T}
\end{equation}
where $\omega= 2 \pi/P$ is the spin rate of the target. For all subsequent computations, we assume that $\rho_{\rm T}/\rho=1$. We also assume, as in \citet{Christou.et.al2017}, that the bulk contribution of impactors comes from Mars-Crossers (MCs) and Near Earth Objects (NEOs) with the distributions of size $s$ and speed $U$ used in that work. Specifically, the speed distribution is extrapolated from the known population of $H\geq18$ impactors while the size distribution is given by a power law, $N(> s) \propto s^{-\alpha}$ \citep{StuartBinzel2004}, where the exponent $\alpha$ is a free parameter in our model.

\begin{figure}
\vspace{-3cm}
\hspace{-5mm}\includegraphics[width=150mm,angle=0]{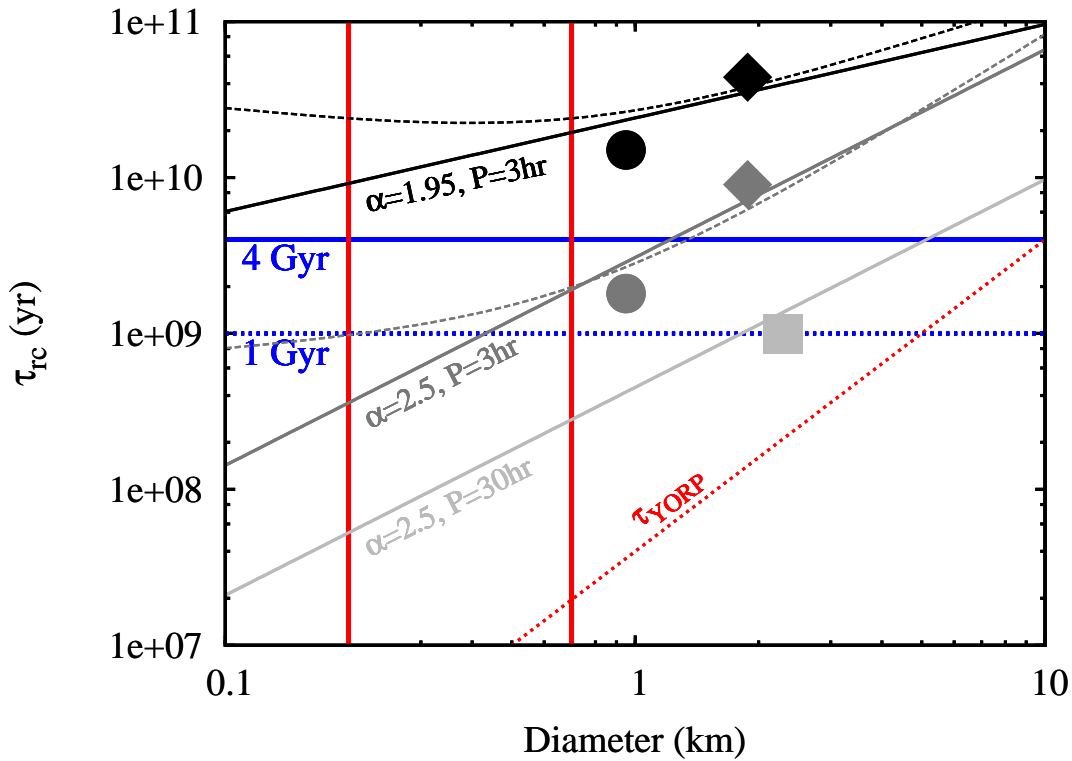}
\caption[Comparison of the collisional spin reset time of the Martian Trojans for different values of the rotation period $P$ and impactor size distribution slope $\alpha$. The diamond, disk and square correspond to Eureka, 101429 and 121514 linked to their respective collisional re-orientation curve - dark-, medium- or light-grey - given their measured or estimated rotation rate and diameter from Table~\ref{tab:troj}. The dashed curves represent the collisional disruption timescale from \citet{Christou.et.al2017} for different values of $\alpha$ and the red dotted line the rotational disruption timescale due to YORP ($\tau_{\rm YORP}$). The vertical solid red lines bracket the likely sizes of known Eureka family asteroids.]{Comparison of the collisional spin reset time of the Martian Trojans for different values of the rotation period $P$ and impactor size distribution slope $\alpha$. The diamond, disk and square correspond to the locations of Eureka, 101429 and 121514 linked to their respective collisional re-orientation curve - dark-, medium- or light-grey - given their measured or estimated rotation rate and diameter from Table~\ref{tab:troj}. The dashed curves represent the collisional disruption timescale from \citet{Christou.et.al2017} for different values of $\alpha$ and the red dotted line the rotational disruption timescale due to YORP ($\tau_{YORP}$). The vertical solid red lines bracket the likely sizes of known Eureka family asteroids.}
\label{fig:t_spin_coll}
\end{figure}

The YORP-induced rotational disruption timescale at 1.5 au is $0.04\mbox{ }Y^{-1} D^{2}$ Myr \citep{Jacobson.et.al2014} where $D$ is the asteroid diameter in km and $Y$ is a shape-dependent numerical factor that is positive or negative depending on the spin direction. Here we adopt $|Y|=10^{-3}$ to obtain 
\begin{equation}
\tau_{YORP}=40\mbox{} D^{2}\mbox{ Myr.}
\label{eqn:tau_yorp}
\end{equation}
We have computed the collisional spin reset time $\tau_{rc}$ as a function of target diameter for values of the impactor size distribution slope $\alpha$ between $1.95$ and $2.5$, as previously used in \citet{Christou.et.al2017} to straddle the range of estimates in the literature. Figure~\ref{fig:t_spin_coll} shows $\tau_{rc}$ (black, dark grey and light grey solid lines) for these limiting values of $\alpha$ and for the different asteroid rotation periods from Table~\ref{tab:troj}. This timescale generally decreases for higher $\alpha$ and for a longer rotation period $P$. Higher values of $\alpha$ also steepen the dependence of $\tau_{rc}$ on $D$, because the relative proportion of impactors able to reset the spins progressively rises with decreasing target size as $\alpha$ increases. The red dotted line represents the timescale for YORP-induced disruption (Eq.~\ref{eqn:tau_yorp}). All the asteroids plot well above this line, suggesting that YORP has reset their rotational states at least several times over. The collisional spin reset times for Eureka, 101429 and 121514 - diamond, disk and square respectively - are intimately linked to the adopted values of $\alpha$ as well as $P$ and we use the brightness level to distinguish between the different cases. The location of Eureka and 101429 relative to their respective $\tau_{rc}$ curves indicates that these asteroids have likely not suffered any rotational-state resetting collisions over the age of the solar system while 121514 may have suffered up to one such collision every Gyr. The dashed curve in Fig.~\ref{fig:t_spin_coll} is the Martian Trojan collisional lifetime from \citet{Christou.et.al2017}. It shows that (i) Martian Trojans generally have their spins reset several times before they are destroyed unless they are fast-spinning ($<$3h), and (ii) km-sized or larger Martian Trojans will, in general, not be eliminated by collisions over 4 Gyr while D$=$200 m objects near the observational completeness limit will survive for about a Gyr for the steepest impactor size distribution. For objects of this size or smaller, it is interesting to note that the spin reset timescale is always shorter than the break-up lifetime regardless of rotation rate.

The spin reset timescale for the Trojans is several times longer than that for similar-sized Main Belt asteroids. For instance, a $D=1$ km object with $P=10$ h will suffer a reset every $10^{9}$ yr for $\alpha=2.5$ while in the Main Belt this occurs every $\sim$$2\times 10^{8}$ yr \citep{Farinella.et.al1998}. This reinforces the conclusion of \citet{Christou.et.al2017} that Martian Trojans enjoy a milder collisional environment than Main Belt asteroids, which was based on their calculation of the collisional {\sl disruption} timescale. 

\section{\label{sec:random}Martian Trojans as a random process}
 Our strategy is to model the observed population as a birth-death stochastic process, where ``offspring'' asteroids are continuously being produced but also lost. At time $t$ we observe $x(t)$ family members with some probability $p(x(t))$. The steady state for such a process follows the Poisson distribution (see Appendix~\ref{sec:app1}) with the mathematical expectation 
\begin{equation}
E[x]=\lambda \tau
\label{eq:exp_steadystate}
\end{equation}
 where $\tau$ is the lifetime for each individual offspring and $ \lambda$ is the production rate of new Trojans from each parent body. We aim to constrain $\lambda$ for the different parent asteroids and for different values of $\tau$, given the observations. The lifetime of new Trojans depends on the loss process efficiency; for instance, we see in Fig.~\ref{fig:t_spin_coll} that the collisional lifetime of the smallest known Trojans varies from $10^{7}$ to $10^{9}$ years depending on the impactor size distribution slope $\alpha$ and the rotation period $P$. Apart from collisional elimination examined in \citet{Christou.et.al2017}, Trojans may be lost if their orbits evolve to the point of escaping the Trojan clouds. We quantify the efficiency of the latter process in the next Section.
 
\section{\label{sec:escape}Trojan Escape}
To find out if Trojans efficiently escape and estimate the loss rate, we integrate clones of the three asteroids under Yarkovsky acceleration strengths representative of the range of sizes of family asteroids. For this we have used the {\it MERCURY} package \citep{Chambers1999}, modified to include the along-track component of the diurnal Yarkovsky acceleration \citep{Farinella.et.al1998}. As in \citet{Christou2013}, we sample the acceleration magnitude $\gamma_{Y}$ regularly and uniformly up to some maximum value $\gamma_{Y,\rm max}$. We note that Trojans may also be lost though chaotic evolution of the orbits under planetary perturbations forces alone, however previous simulations \citep{Cuk.et.al2015} show that the rate of orbit dispersion under Yarkovsky is at least as high as that under gravitational forces. In fact, in our simulations we observe that it is typically the smallest Trojans, i.e.~those that evolve under the highest values of $\gamma_{Y}$, that escape first.
The solar system model consisted of the eight major planets from Mercury to Neptune, integrated with a time step of $4$ days and an output step of $10^{5}$ yr. We used initial planetary state vectors at JD2451545.0 recovered from JPL HORIZONS \citep{Giorgini.et.al1996} while for the three Trojans we used HORIZONS state vectors at JD2454613.5 (Eureka), JD2455169.5 (101429) and JD2455138.5 (121514). Clones were produced from these nominal orbits by sampling the state covariance of-date, also available from HORIZONS and the clone orbits propagated to the same epoch as the planets at the beginning of the runs.

As we are interested to know when the Trojans escape, we chose to use the ``hybrid'' symplectic scheme available within {\it MERCURY}. The scheme accurately handles close encounters between particles and planets by switching from mixed-variable symplectic to Bulirsch-Stoer state propagation within a certain distance from a planet. For all the simulations reported here, this changeover threshold was set at 2 Hill radii.

\begin{figure}
\vspace{-3cm}
\hspace{-2cm}\includegraphics[width=90mm,angle=0]{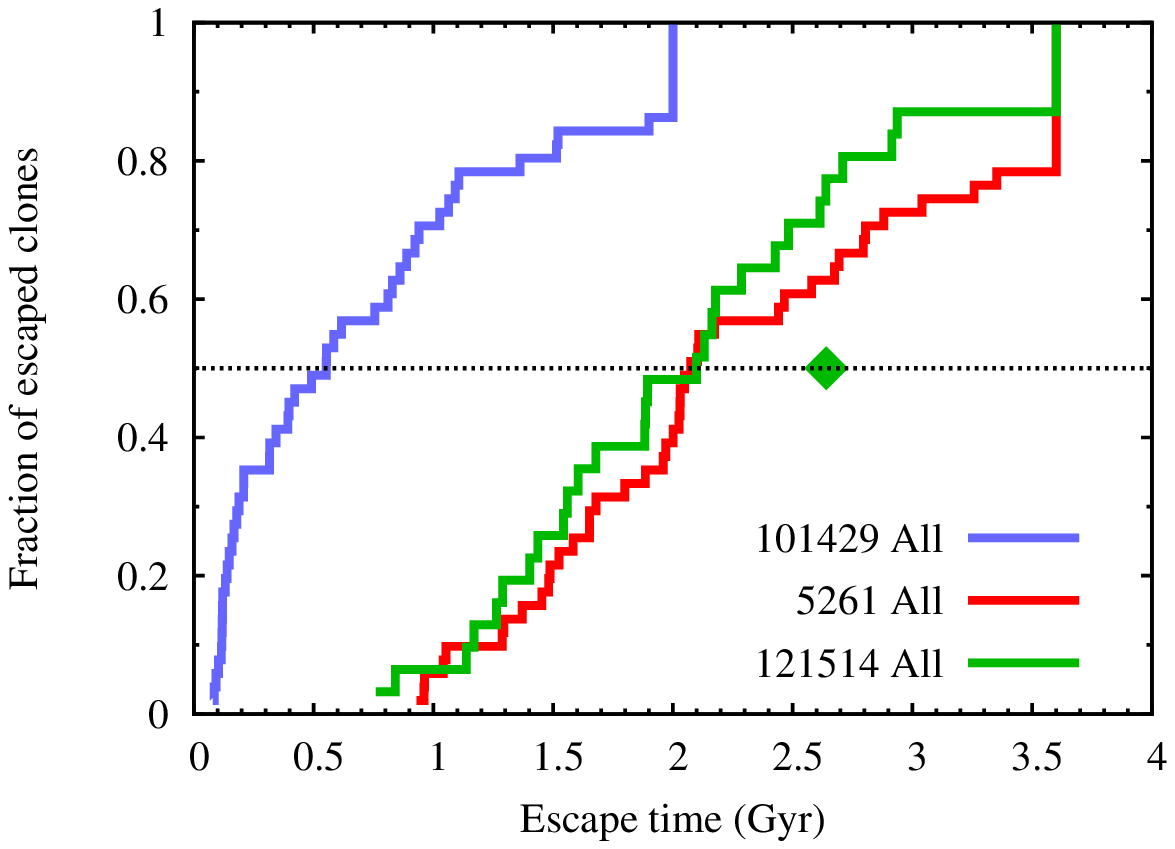}\includegraphics[width=90mm,angle=0]{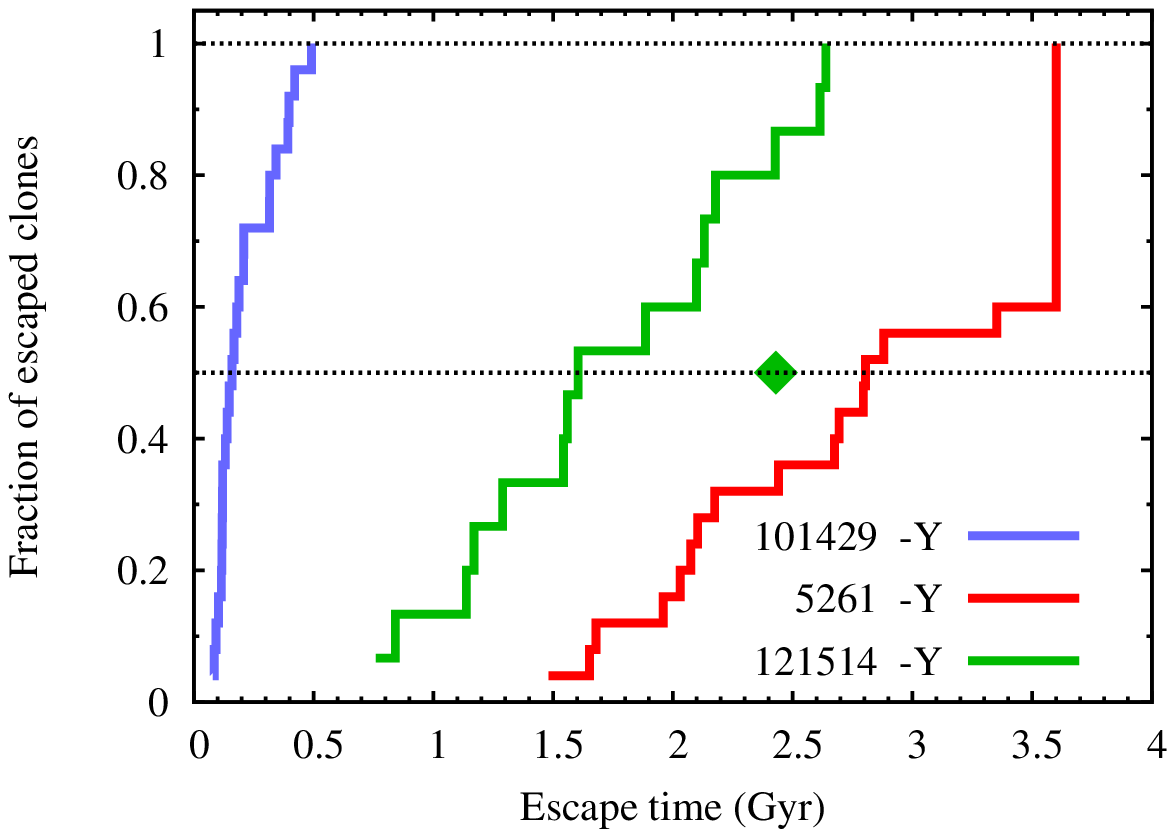}
\caption[Cumulative distribution of escape time for Martian Trojans originating from Eureka (red), 101429 (blue) and 121514 (green) in our numerical runs. Each distribution is terminated by a vertical line to indicate the end of the run for each object (see Table~\ref{tab:sims}). Left: All Trojan clones. Right: Only those clones evolving under negative Yarkovsky acceleration. The diamond represents the half-life of family members of 121514 if we adopt a higher limiting size than for the other two Trojans due to the lower albedo of this asteroid.]{Cumulative distribution of escape time for Martian Trojans originating from Eureka (red), 101429 (blue) and 121514 (green) in our numerical simulations. Each distribution is terminated by a vertical line to indicate the end of the run for each object (see Table~\ref{tab:sims}). Left: All Trojan clones. Right: Only those clones evolving under negative Yarkovsky acceleration. The diamond represents the half-life of family members of 121514 if we adopt a higher limiting size than for the other two Trojans due to the lower albedo of this asteroid.}
\label{fig:escapes}
\end{figure}

Our simulations are summarised in Table~\ref{tab:sims}. Our chosen value of $\gamma_{Y,\rm max}$ for all three Trojans is approximately that of a $D=300$ m asteroid of visual albedo $p_{\rm v}=0.2$ for which $H=20$. A consequence of limiting the maximum Yarkovsky acceleration in terms of asteroid size rather than the brightness is that, due to the $4-5\times$ lower albedo of 121514 compared to the other two asteroids, we will be underestimating the physical sizes of the smallest observable offspring from this object, by a factor of $\sim$2; all else being equal, this will overestimate the acceleration strength by a similar factor since $\gamma_{Y}$$\propto$$D^{-1}$. In spite of this, we do not see an advantage in using the higher limiting size for asteroids within a putative 121514 family in this work, for two main reasons. Firstly, it only has a moderate effect on the estimated lifetimes (green diamonds vs green curve in Fig.~\ref{fig:escapes}); the value of $\tau_{121514}$ increases by $0.5$ Gyr (a fractional change of 25\%) in the case of sign-independent Yarkovsky acceleration and by $0.8$ Gyr (fractional change of 50\%) for a negative-only acceleration. Secondly, the value of $\gamma_{Y,\rm max}$ depends, apart from the albedo, on poorly constrained properties such as bulk and surface density and the thermal conductivity, the combined effect of which can be on a par with that of the albedo. In fact, the constraints we are able to impose on the production rate of offspring asteroids (Section~\ref{sec:constraints}) are insensitive to $\gamma_{Y,\rm max}$, depending instead on the stability properties of the phase space occupied by the Martian Trojans.

Overall, we observe that escape of asteroids from the Martian Trojan clouds is prevalent in our simulations. The distribution of escape times for the Trojans is shown in the left panel of Fig.~\ref{fig:escapes}, colour-coded to distinguish between the different parent bodies. In the right panel we show the escape statistics only for those asteroids evolving under a negative Yarkovsky acceleration as per \citet{Cuk.et.al2015}. Vertical line segments, where present on the right end of each distribution, provide a visual indication of the surviving fraction of Trojans from each parent asteroid. For instance, in the right-hand panel we see that $\simeq$40\% of offspring from Eureka evolving under a negative Yarkovsky acceleration survived the full 3.6 Gyr.
     
It is immediately clear that offspring from 101429 escape much faster than for the other two parent bodies. They begin escaping $\sim$100 Myr into the simulations (left panel) with 80\% of their initial set of clones lost from the Trojan clouds after 2 Gyr. Clones of that asteroid evolving under negative $\gamma_{Y}$ (right panel) escape first with no such clones remaining in the simulation after 500 Myr. In contrast, the first escapes of offspring from Eureka and 121514 occur at $\sim$800 Myr while, under negative Yarkovsky, clones of Eureka begin to escape later, at 1.5 Gyr. We also note the dissimilar shapes of the distributions, with those for 5261 and 121514 appearing linear-like, while that for 101429 being more similar to a decay function.   

The short lifetime of asteroids from 101429 is probably related to its location near the $\nu_{5}$ secular resonance at $30^{\circ}$ inclination \citep{Scholl.et.al2005}, slightly below the asteroid's $I \simeq 32^{\circ}$ \citep{Christou2013}. The resonance may act as an efficient ``escape hatch'', similar to the role played by e.g.~the $\nu_{6}$ resonance and mean motion resonances with Jupiter in the Main Belt \citep{Gladman.et.al1997}. It also raises the question of the dynamical lifetime of 101429 itself. To answer this, we have generated 20 additional clones of this asteroid and integrated them for 1 Gyr with the Yarkovsky acceleration feature turned off. None of these clones escaped, which suggests that Yarkovsky plays a crucial role in the long-term stability of this object. 
On the other hand, the Yarkovsky clones of 101429 with the smallest $|\dot{a}|$ -- $4 \times 10^{-5}$ au $\mbox{Myr}^{-1}$ and 6$\times$ lower than the nominal value for a 1-km asteroid from our force model -- all escaped; those clones that survived for 2 Gyr have $\dot{a}$ in the range +($1.0$--$1.5$)$\times 10^{-3}$ au $\mbox{Myr}^{-1}$. Because the lifetime of clones depends on the acceleration magnitude in a non-simple way, we speculate that this behaviour is again related to the proximity of 101429 to the secular resonance and that the boundary between the stable domain of orbit space, where 101429 is presently located, and the region of instability owing to the resonance may be topologically complex. This conclusion has implications for the assumptions under which this study was conducted and we shall return to it in Section~\ref{sec:end}.

We now carry on to the original aim of this exercise, namely to constrain the dynamical lifetimes of Trojans under Yarkovsky. For simplicity, we would like to represent this with a single number for each object and we choose for this purpose the time at which 50\% of the clones have escaped. For 101429 and 5261 this corresponds to the 26th shortest escape time or the 13th shortest 
 among the 25 clones with $\gamma_{Y}<0$. For 121514 we use the 16th and 8th shortest escape time respectively due to the smaller number of clones used.  For 101429 this yields, rounding to the nearest 50 Myr, $\tau_{101429}=0.55$ Gyr in the first instance or $\tau_{101429}=0.15$ Gyr if we only consider clones with $\gamma_{Y}<0$. For Eureka and 121514 we respectively obtain $\tau_{Eureka}=2.05$ Gyr and $\tau_{121514}=2.10$ Gyr, or $\tau_{Eureka}=2.90$ Gyr and $\tau_{121514}=1.60$ Gyr for the subset with $\gamma_{Y}<0$.  Interestingly, clones of 121514 with $\gamma_{Y}<0$ typically escape earlier than the remaining clones for the same object, while for Eureka the opposite is true.

Finally, we compare our long-term orbit simulations of Martian Trojans with those of \citet{Cuk.et.al2015} for consistency and find (Appendix~\ref{sec:app2}) that the two Yarkovsky implementations are effectively equivalent in that they produce similar outcomes from similar inputs.

\section{\label{sec:existence}Defining the observable: does non-detection imply non-existence?}
Before we can apply the steady state model to the Martian Trojans, we must grapple with the fact that both 101429 and 121514 are intrinsically fainter than Eureka, albeit for different reasons: 101429 is physically smaller than Eureka whereas 121514 is darker. If offspring asteroids related to these two objects exist and their sizes relative to the parent bodies are Eureka-like, they will have $H\gtrsim17+2 =19$ and as faint as $\sim 20$ if they are similar to Main Belt asteroid clusters \citep{Pravec.et.al2018}. Therefore, we need to find out if, or to what degree,  their non-detection up to now is merely the result of observational incompleteness.   This is done here by generating synthetic families of Martian Trojans and simulating their detection by past and ongoing asteroid surveys in a Monte Carlo model. In this task we are aided by the availability of a proxy population, namely Main Belt asteroid clusters \citep{Pravec.et.al2018} to help constrain the expected population characteristics of YORP fission products.

\subsection{\label{sec:synthetic}Generating synthetic families}
\begin{figure}
\vspace{-3cm}
\hspace{-2cm}\includegraphics[width=90mm,angle=0]{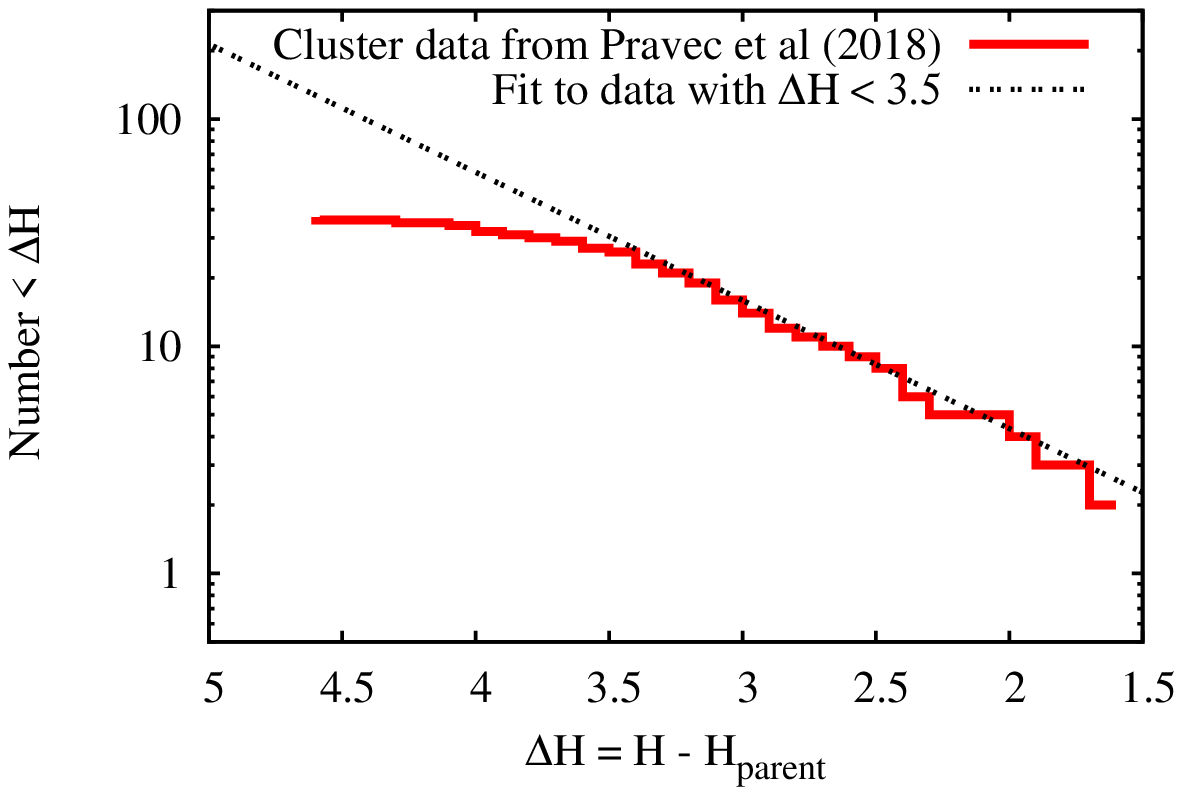}\includegraphics[width=90mm,angle=0]{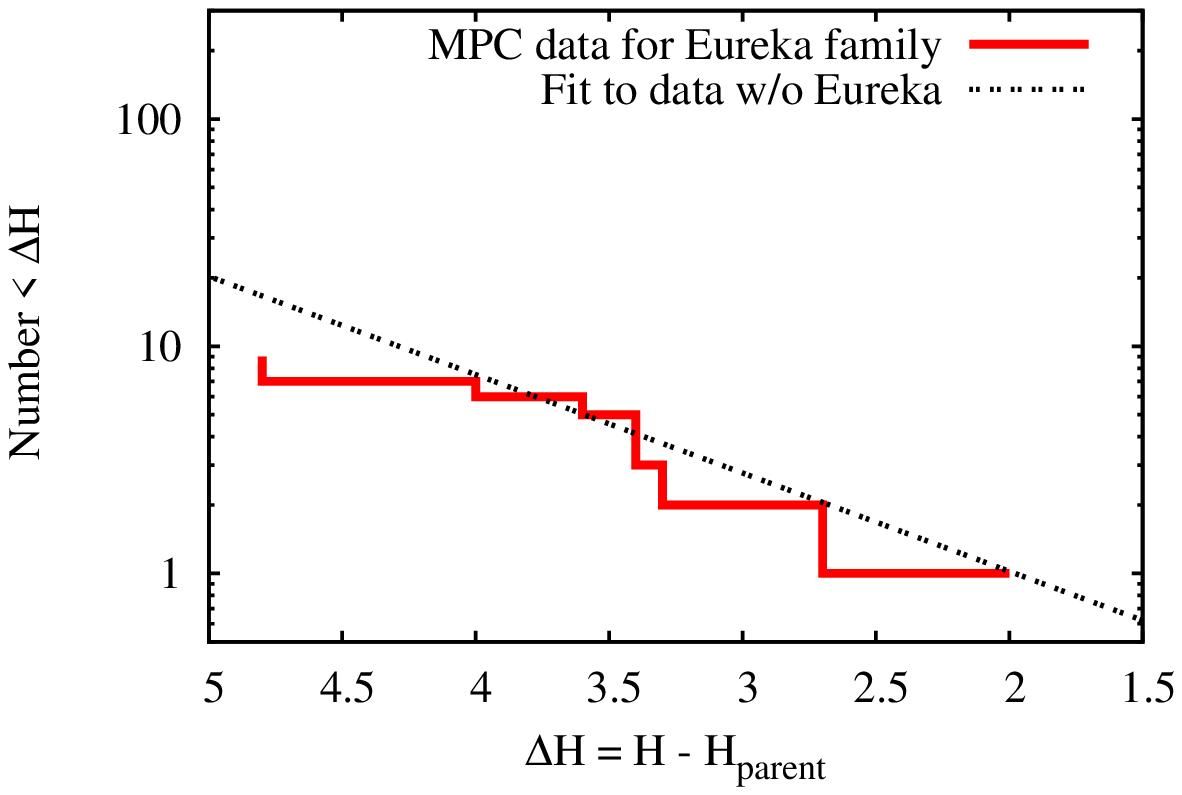}
\caption[Left: Cumulative distribution (red line) of relative absolute magnitudes for cluster asteroids listed in \citet{Pravec.et.al2018}. The dashed line represents a power law fit to the data with $\Delta H<3.5$. Right: Cumulative distribution of absolute magnitudes for Eureka family asteroids. The dashed line represents a power law fit to the data up to $\Delta H=4$.]{Left: Cumulative distribution (red line) of relative absolute magnitudes for cluster asteroids listed in \citet{Pravec.et.al2018}. The dashed line represents a power law fit to the data with $\Delta H<3.5$. Right: Cumulative distribution of absolute magnitudes for Eureka family asteroids (red line). The dashed line represents a power law fit to the data up to $\Delta H=4$.}
\label{fig:clusters}
\end{figure}

We find it convenient to express the absolute magnitude of YORPlets as
\begin{equation}
\label{eqn:absmag}
H_{k}=H_0+\Delta H_{k}
\end{equation}
where $H_{0}$ is the magnitude of the primary and $\Delta H_{k}$ is the difference between the parent body and the $k$th cluster member in order of increasing $H$. For the Eureka family, $\Delta H_{1}=2.0$. \citet{Pravec.et.al2018} report 11 clusters of small (12$<$$H$$<$19) Main Belt asteroids that may be the product of one or more episodes of rotational fission. We have calculated values of $\Delta H_{k}$ for 9 of these clusters. We have not included (a) two clusters at odds with predictions from rotational fission theory that may have formed through a different mechanism \citep{Pravec.et.al2018} and (b) three individual asteroids with one-opposition orbits. The cumulative distribution of $\Delta H$ for the remaining 36 asteroids is shown in the left panel of Fig.~\ref{fig:clusters}. The rolloff for $ \Delta H \lesssim 3.5$ is probably due to sample incompleteness for the fainter asteroids. We perform a linear regression fit to a power law of the form $\log_{10} N= a + b H$ to the asteroids with $\Delta H \leq 3.5$ and find $a=-0.49 \pm 0.05$ and $b=0.56 \pm 0.02$ (dashed-dotted line in Fig.~\ref{fig:clusters}). 

If we carry out the same exercise for seven members of the Eureka family\footnote{we do not include 2011 $\mbox{SP}_{189}$ as it is significantly fainter than the other members and doing so will probably skew the slope estimate away from the true value} we find $a= -0.89 \pm 0.15$ \& $b = 0. 45 \pm 0.05$ (right panel of Fig.~\ref{fig:clusters}). Note that the slope is steeper than found in a fit to the same data in \citet{Christou.et.al2017}. This is because that other fit also included the parent body - Eureka - and was done to compare with the overall size distribution of small Main Belt asteroids. Here, our use of specific asteroid clusters allows us to separate out the parent bodies before carrying out the fit.

\begin{figure}
\vspace{-3cm}
\includegraphics[width=85mm,angle=0]{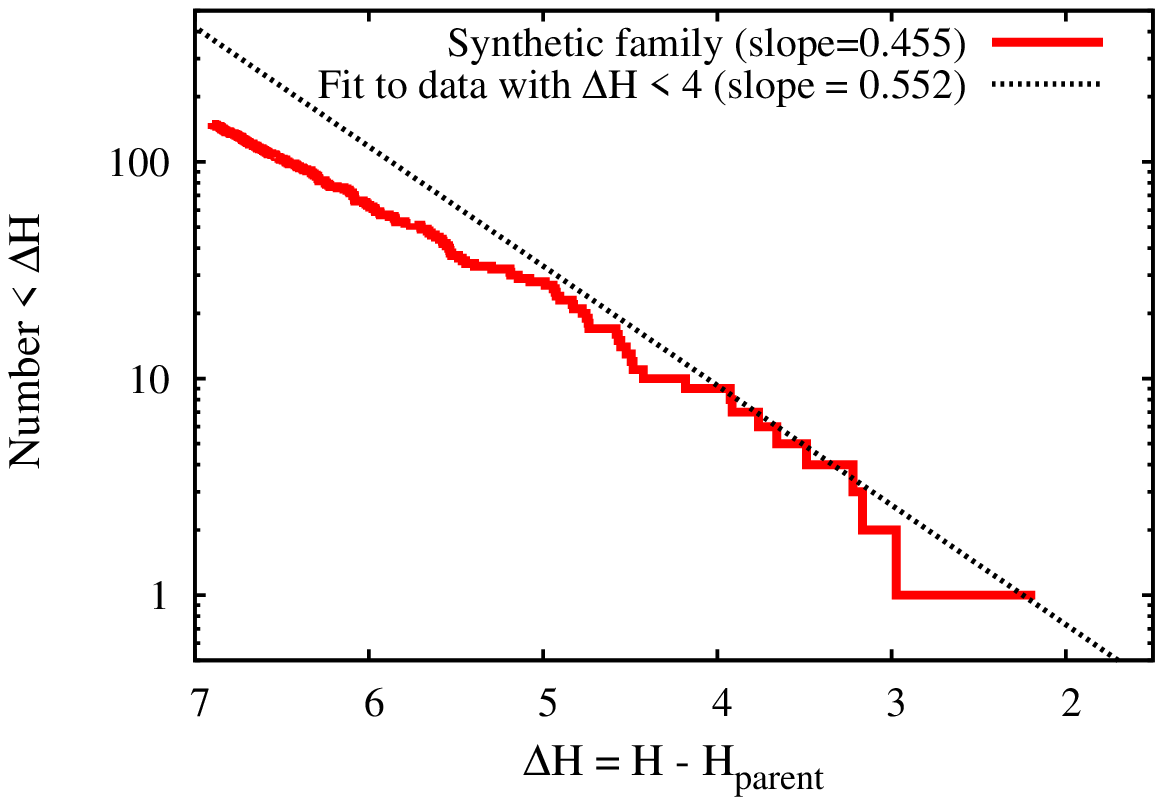}\includegraphics[width=80mm,angle=0]{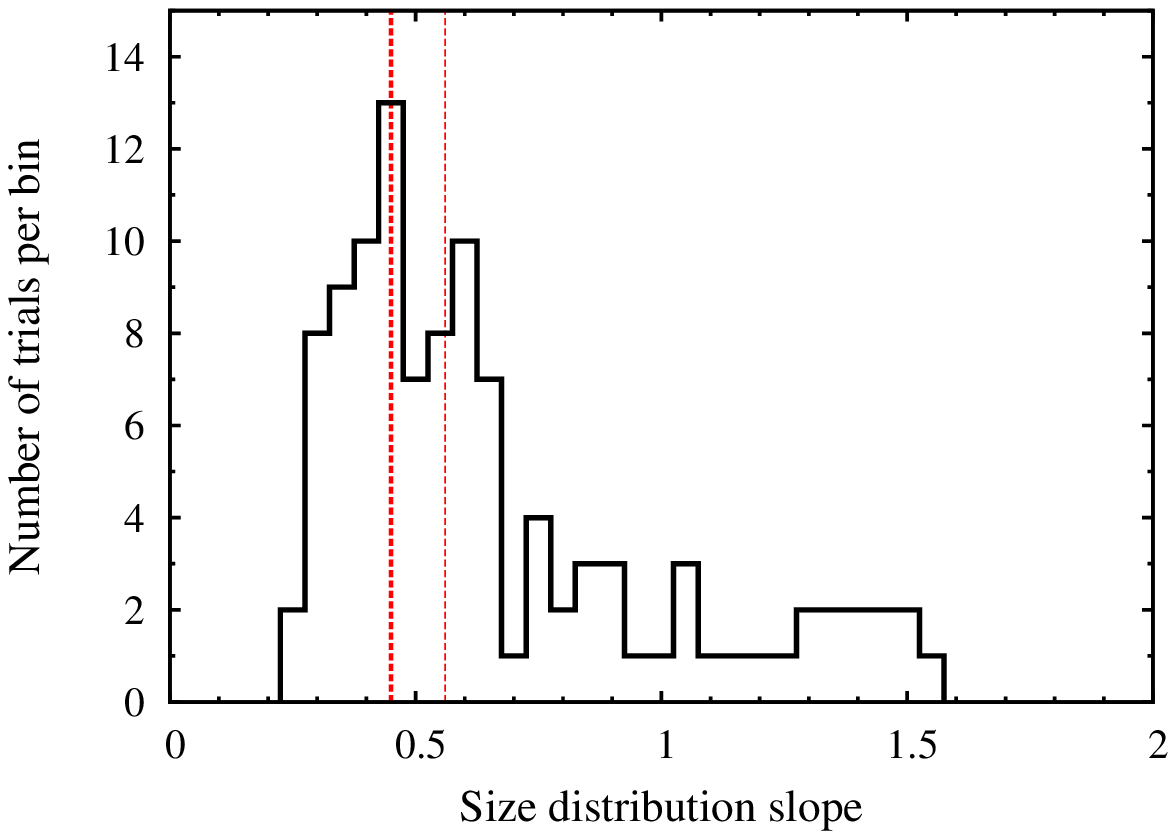}
\caption[ Left: As Fig.~\ref{fig:clusters}, left panel, but for a numerically generated synthetic family. The dashed line represents a power law fit to the data with $\Delta H<4$. Right: Distribution of best-fit slopes in 100 randomly-generated synthetic families with $b=0.45$ and $N(\Delta H < 4)=7$. The dashed vertical lines correspond to the values $b=0.45$ (thick line) and $b=0.56$ (thin line) considered in the Monte Carlo runs.]{Left: As Fig.~\ref{fig:clusters}, left panel, but for a numerically generated synthetic family. The dashed line represents a power law fit to the data with $\Delta H<4$. Right: Distribution of best-fit slopes in 100 randomly-generated synthetic families with $b=0.45$ and $N(\Delta H < 4)=7$. The dashed vertical lines correspond to the values $b=0.45$ (thick line) and $b=0.56$ (thin line) considered in the Monte Carlo runs.}
\label{fig:synthetic}
\end{figure}

To generate synthetic families with the prescribed magnitude distribution, we randomly sample the power-law magnitude distribution function in terms of the magnitude of the largest member $H_{1}$, the slope $b$ and the faintest magnitude $H_{max}$ of cluster members to be generated. The distribution is scaled by requiring that $N(H_{1})=1$. Figure~\ref{fig:synthetic}
shows an example of a family of 151 asteroids generated using the fit parameters obtained for the Eureka family i.e. with $b = 0. 45$. Each sampling of the distribution constitutes a trial in the Monte Carlo runs. Note that, if we carry out a new fit to the synthetic data for $\Delta H <4$ - a slightly fainter limit than the 3.5 used in the left panel of Fig.~\ref{fig:clusters} in order to allow a good fit - we find $b=0.55\pm 0.03$, very similar to the fit for the \citeauthor{Pravec.et.al2018} data. To investigate further, we fit power-law slopes to 100 randomly generated families with $b=0.45$ and $N(\Delta H < 4)= 7$. The resulting distribution of slope values is shown in Figure~\ref{fig:synthetic}, right panel. The tail seen for values $b>0.7$ and as high as $1.5$ corresponds to families where the brightest randomly-generated member is significantly fainter than $H=18.1$, producing a steeper slope.  The values obtained from the fit to the actual Eureka family ($b=0.45$) and the asteroid cluster data ($b=0.56$) are represented by the thick red and thin red lines respectively. Both values plot near the mode of the distribution, serving to emphasise that the gross properties of this family are, at the moment, statistically indistinguishable from a typical asteroid cluster produced by rotational fission. 

\subsection{\label{sec:survey}Simulating Martian Trojan discovery in asteroid surveys}
Contrary to what is common practice in survey simulations \cite[eg ][]{Grav.et.al2016,SchunovaLilly.et.al2017}, here we do not implement a particular survey strategy in the form of specific observation epochs or instrument pointings. We feel justified in doing so because the objects of interest - families of Martian Trojan asteroids - move in very similar orbits and occupy a compact subset of those orbits targeted by large-scale solar system surveys. Therefore, the number of detections will scale principally with the amount of time that the object's magnitude is above the detection threshold of the instrument rather than the orbit.
A disadvantage of taking this approach is the need to devise an alternative method to calibrate the detection
efficiency of our ``pseudo-survey''. We do this by utilising the available knowledge on the Eureka family and by exploiting the fact that the survey sensitivity can be tied to the absolute magnitude $H$ of the target asteroids in a unique way.

In our pseudo-survey we assume that all family asteroids move along fixed keplerian orbits with the same semimajor axis, eccentricity and inclination as the candidate parent body (Eureka, 101429 or 121514) at JD2451545.0, retrieved from HORIZONS. For a given epoch, the state vectors of the Earth and the asteroid relative to the Sun are calculated in the usual way by solving Kepler's equation \citep{MurrayDermott1999} and the mean anomaly propagated from the ephemeris value for the parent asteroid at the reference epoch. 
In a survey field containing the sky location of the asteroid, the probability of detection at that epoch is given by 
\begin{equation}
p(V)=p(V; V_{50}, w)= {\left( 1 + e^{\frac{(V-V_{50})} {w}}\right)}^{-1}
\label{eqn:p_detect}
\end{equation}
where $V_{50}$ is the value of $V$ for which this probability is 50\%. Here we adopt the values $V_{50}=21.0$ and $w=0.2$  applicable to the two most sensitive asteroid surveys considered by \citet{Tricarico2016}: Pan-STARRS (Obs.~code F51) and the Mt Lemmon Survey (Obs.~code G96).

The magnitude $V$ for a given Earth-Sun-asteroid geometry is calculated as \citep{Bowell.et.al1989}
\begin{equation}
V=H-5 \log{(r \Delta)} - 2.5 \log [(1 - G) \Phi_{1}(\phi) +  G \Phi_{2}(\phi)]
\label{eqn:phase}
\end{equation}
where $r$ \& $\Delta$ are the heliocentric and geocentric distances of the asteroid respectively, $\Phi_{1}$ and $\Phi_{2}$
are known functions of the phase angle $\phi$ and $G$ is the slope parameter, set here to $0.15$. 

To decide whether a detection has taken place or not, we generate a random number $x$ between 0 and 1. A source is {\it detected} if $0<x<p(V)$ {\it and} the solar elongation $E$ of the target is higher than some limit $E_{\rm min}$, and {\it not detected} if either $p(V)<x<1$ {\it or} $E<E_{\rm min}$. This is done for a sample of $N$ regularly spaced epochs within an interval defined by a start epoch  $T_{0}$ and an end epoch $T_{1}$. 

The detectability of moving objects in a sky survey also depends on so-called ``trailing losses'', the smearing out of the signal from the source across several pixels on the CCD frame. \citet{HarrisDAbramo2015} found a significant bias against the discovery of faint new NEAs if the rate of motion is in excess of 2-3 deg/day.  To determine if this is a concern for Martian Trojans, we have used the HORIZONS ephemeris service to tabulate the geocentric apparent rate of motion every 10 days during the 17-year period 1-Jan-2003 - 31-Dec-2019 for all Trojans in Table~\ref{tab:troj}. We find that the rate of motion is always $\lesssim 1$ deg/day. Therefore, trailing losses are probably not important for Martian Trojans and we opt not to model their effects here.

The survey parameters used in the Monte Carlo simulations are listed in Table~\ref{tab:sims2}. The adopted value for $E_{\rm min}$ was chosen as one through which the efficiency of detection with a ground-based facility drops rapidly in the direction approaching the Sun. Here we have used as a proxy the projected Large Synoptic Survey Telescope performance for NEA detection from \citet{ChesleyVerres2017}. This choice is not unique and a slightly different value e.g.~$80^{\circ}$ could be chosen, yet the sensitivity of the survey simulation to this parameter is low because Martian Trojans near the elongation limit are relatively far from the Earth and less likely to contribute to the discovery statistics. We choose to run our pseudo-survey for a 20-year period starting in June 1998 to cover the interval of time during which all of the presently known Trojans, apart from Eureka, were discovered. These choices ensure that the objects are observed under a wide range of observational geometries including during the most favourable oppositions which, like for Mars itself, recur every 15 to 17 yr.

\begin{figure*}
\vspace{-3cm}
\centering
\includegraphics[width=80mm,angle=0]{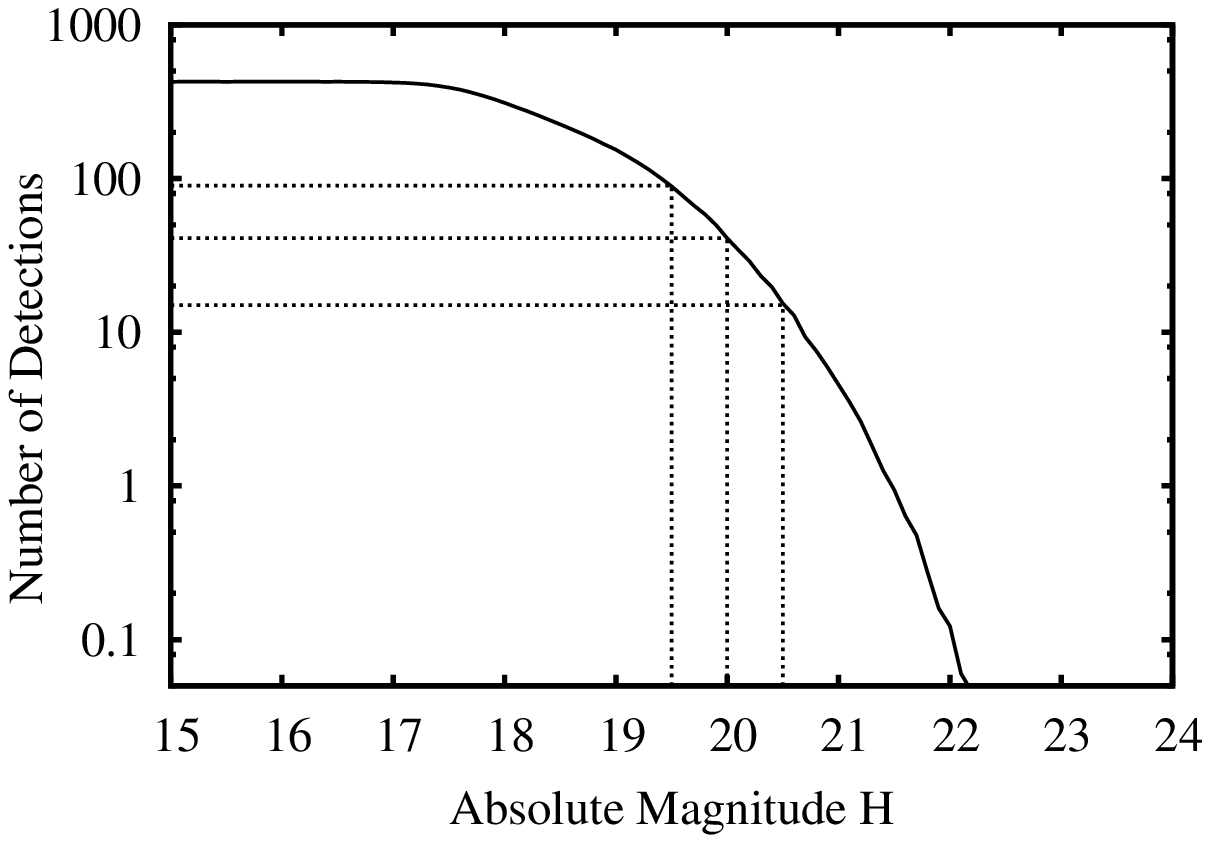}\includegraphics[width=80mm,angle=0]{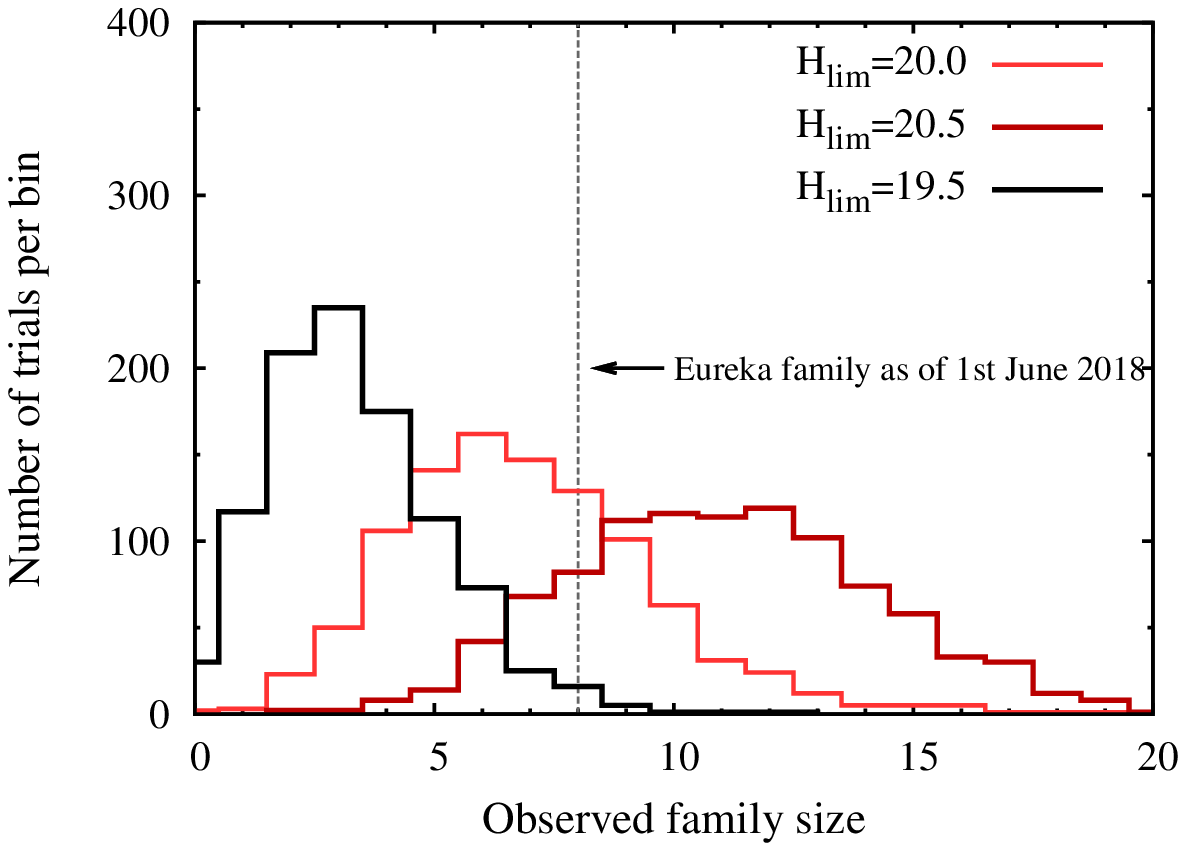}
\caption[Left: Number of detections for Mars Trojans in Eureka-like orbits and varying $H$. The dashed vertical lines indicate different values for the number of detections $n_{d}$ used to calibrate survey completeness. Right: Distributions of different outcomes from $N_{t}=1000$ randomly-generated Eureka-like families. The distribution for $H_{lim}$=20 (light red) is the most likely to reproduce the observed size of the Eureka family.]{Left:  Number of detections for Mars Trojans in Eureka-like orbits and varying $H$. The dashed vertical lines indicate different values for the number of detections $n_{d}$ used to calibrate survey completeness. Right: Distributions of outcomes from $N_{t}=1000$ randomly-generated Eureka-like families. The distribution for $H_{lim}$=20 (light red) is the most likely to reproduce the observed size of the Eureka family.}
\label{fig:detections}
\end{figure*}

We utilise our algorithm in two stages. In the first, `calibration', stage we initially simulate the detection of single asteroids in Eureka-like orbits with absolute magnitude $H$ in the range 16-24 and in steps of 0.1. For each value of $H$ we generate $N^{2}_{\rm o}$ test orbits with spatial orientation parameterised by the argument of perihelion and longitude of ascending node, drawing $N_{\rm o}$ random variates for each angle between $0$ and $2 \pi$. The chosen value $N_{\rm o}=30$ ensures a good statistical sampling of the different orbit orientations with a modest computing time penalty. The number of observation epochs is set to $N_{e}=1000$. The other orbital elements are fixed to their reference values for Eureka with the mean anomaly propagated from the value at the reference epoch. The result (left panel of Fig.~\ref{fig:detections}) shows a clear relationship between the number of detections $n_{d}$ over the $N_{e}$ epochs - averaged over the different orbit orientations by dividing by $N^{2}_{\rm o}$ - and the absolute magnitude $H$. Therefore, $n_{d}$ and $H$ are equivalent quantities and may be used interchangeably. In particular, the effective limiting magnitude $H_{lim}$ of our pseudo-survey can be derived by finding a threshold value for the number of detections  that defines the survey efficiency. 

With this in mind, we run $N_{t}=1000$ Monte Carlo trials of our survey simulator, each time generating a new synthetic family of asteroids down to $H_{max}=23$ with the same magnitude distribution as found from our power-law fit to the Eureka family i.e.~with slope $b=0.45$ and $H_{1}=18.1$. We repeat this exercise for different choices of the threshold number of detections, equivalent to the following values of $H_{lim}$: 19.5, 20.0 and 20.5 (Fig.~\ref{fig:detections}, left panel). The other control parameters of our survey simulation remain the same. We find (Fig.~\ref{fig:detections}, right panel) that, for the two latter values of $H_{lim}$, the observed size of the family (8 asteroids, not counting Eureka) plots near the mode of the distribution and has a relatively high likelihood of being obtained from the survey, higher under the assumption $H_{lim}=20.0$ (bright red histogram) than for $H_{lim}=20.5$ (dark red histogram). Although in principle one can continue improving the formal estimate of $H_{lim}$ to $<0.5^{\rm m}$, given that (i) in our survey simulations we are not utilising actual pointing histories, and (ii) the observed population size of eight precludes using a more significant observable (e.g.~the magnitude distribution) at this time, we choose to stop the procedure here and adopt the value $H_{lim}=20$ which, from the left panel of Fig.~\ref{fig:detections}, corresponds to $n_{d}=41$ detections. 
 
In deciding what value of $H_{lim}$ - and therefore the $n_{d}$ threshold - may apply to our survey, our reasoning was as follows: Ideally, the limiting absolute magnitude should correspond to 50\% completeness, i.e.~about half of all asteroids brighter than this have been discovered. It follows that this magnitude should be brighter than the faintest member of the Eureka family, therefore $H_{lim}<21$. Indeed, the fact that all family members apart from 2011 $\mbox{SP}_{189}$ have $H \lesssim 20$ or brighter, suggests that the actual limiting magnitude is significantly brighter than 21. At the same time, we note that 311999 and 385250, the two brightest family members other than Eureka itself, have H$<$19. Those asteroids were recognised as stable Martian Trojans more than a decade ago and it seems unlikely that two - or even one - Trojans with $H\lesssim19$ remain undiscovered to this day. These two arguments combined show that $19 < H_{lim} < 21$.

Having obtained the threshold value of $n_{d}$, we adopt it in the second, `survey', stage of our Monte Carlo runs. In other words, any asteroid that is detected at least 41 times in survey runs with $N_{e}=1000$ is ``discovered'' as a Martian Trojan. For each randomly generated test orbit we again choose $N_{e}$ observation epochs but now randomly drawn between $T_{0}$ and $T_{1}$.  We carry out $N_{t}=1000$ Monte Carlo trials as before but with $N_{e}=100$ instead of $N_{e}=1000$ or one observation every 70 days and by sampling the orientation of the target orbit only once, that is $N_{o}\times N_{o}=1$. We do this to speed up the computations, allowing to generate each of the distributions shown in Fig.~\ref{fig:detections} in $\sim$15 min on a portable computer. Since we are using a sparser set of simulated observations, we scale our detection criterion accordingly and require at least [41/10]=4 detections to record a test object as `discovered'.  We emphasise that this number should not be viewed as the number of times an asteroid needs to be observed to be confirmed as a Eureka family member. Rather, it is tied to the specific choices of survey parameter values shown in Table~\ref{tab:sims2}. 

The results of our Monte Carlo runs are parameterised in terms of the slope $b$ of the magnitude distribution and the difference in magnitude $\Delta H_{1}$ between the parent body and the brightest member. The values of $\Delta H_{1}$ for the fission clusters reported in \citet{Pravec.et.al2018} are, in increasing order: 1.6, 1.7, 1.7, 2.0, 2.4, 2.6, 2.6, 2.7 \& 2.9 (their Table~1). We choose to sample $\Delta H_{1}$ between the values $1.5$ and $3.0$ in increments of 0.5.

\subsection{\label{sec:results}Simulation results}
\begin{figure}
\vspace{-3cm}
\centering
\includegraphics[width=70mm,angle=0]{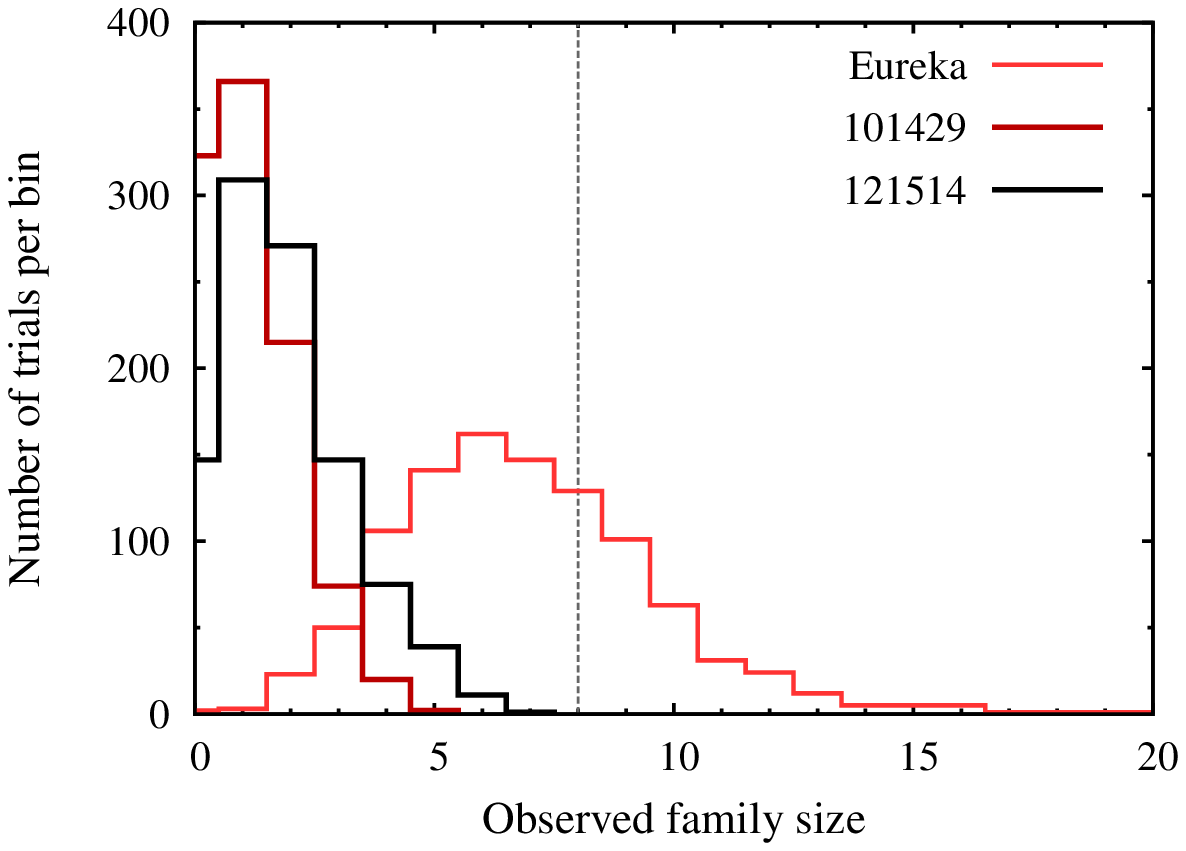}\includegraphics[width=70mm,angle=0]{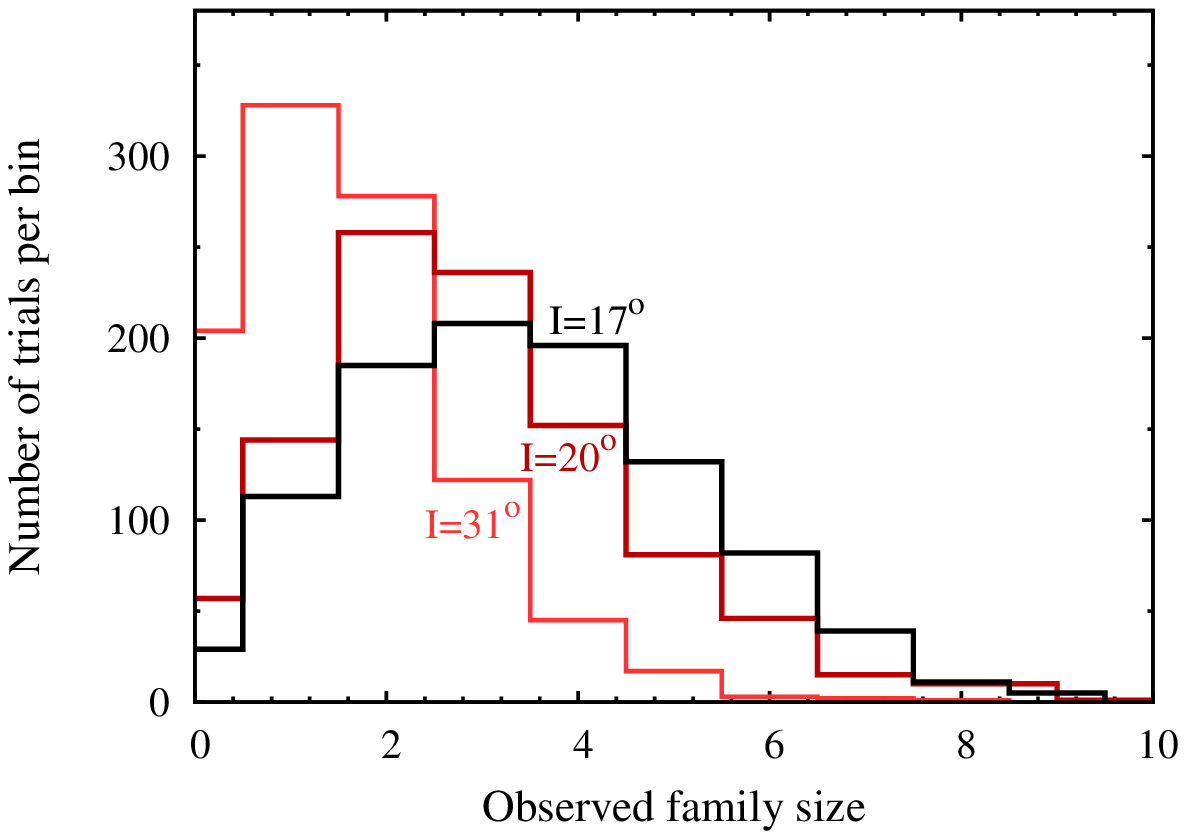}
\caption[Left: Family size statistics from the Monte Carlo simulations and for different parent bodies. The dashed vertical line indicates the current number of asteroids in the Eureka family. Right: Statistics for asteroids associated with 101429 for three different values of the orbital inclination.]{Left: Family size statistics from the Monte Carlo simulations and for different parent bodies. The dashed vertical line indicates the current number of asteroids in the Eureka family. Right: Statistics for asteroids associated with 101429 for three different values of the orbital inclination.}
\label{fig:mc_example_inc}
\end{figure}

The result of each run is a distribution of absolute magnitudes for the detected family asteroids. An example is shown in Fig.~\ref{fig:mc_example_inc}, left panel, where we have set ${\Delta H}_{1}=2$ for all three objects. We note several features of interest here. The observed size of the Eureka family is near the mode of the distribution for Eureka.  This may be seen as a confidence-building check of the `survey' version of the code, given that it uses a 10-fold sparser set of observations than the `calibration' version. Also, we apparently detect fewer family members of 121514 and 101429 than for Eureka, simply because these asteroids are fainter, by $\sim1.0$ magnitude. In particular, non-detection of an extant family is twice more likely for 101429 than for 121514. We have investigated this further and found it is related to 101429's higher orbital inclination, $\sim$$30^{\circ}$ compared to $17^{\circ}$ for 121514. We demonstrate this in the right panel of Fig.~\ref{fig:mc_example_inc} where we have used 101429's orbit and assigned, for each run, an inclination from the orbit of each of the three potential parent bodies. At $I=17^{\circ}$ only $\sim$30/1000 trials yield zero detections, compared to 200/1000 for $I=31^{\circ}$, because asteroids at a higher inclination orbit will spend more time far from the ecliptic plane - and therefore from the observer - than at lower inclinations. This bias is mitigated somewhat by 101429's slightly higher orbital eccentricity but, in any case, it means that, at the current level of observational completeness we are less certain of the non-existence of a family associated with this asteroid than for 121514.  

The Monte Carlo runs yield an estimate of the probability $p(\mbox{no detections; }\overrightarrow{\mathit{\Delta} H}_{1})$ for different values of the parameter vector $\overrightarrow{\mathit{\Delta} H}_{1}=\left(\mathit{\Delta} H^{101429}_{1}\mbox{, }\mathit{\Delta} H^{121514}_{1}\right)$. This is effectively a likelihood function for $\overrightarrow{\mathit{\Delta} H}_{1}$ and can be readily converted to an {\it a posteriori} probability distribution by multiplying with $p(\overrightarrow{\mathit{\Delta} H}_{1}$), the {\it a priori} distribution of the parameters. If we assume the latter distribution to be uniform, we obtain, after summing and normalising, the  {\it a posteriori} cumulative distribution $p(\overrightarrow{\mathit{\Delta} H}_{1} \mbox{; no detections})$ in Table~\ref{tab:sign_0p44} for $b=0.45$ and in Table~\ref{tab:sign_0p56} for $b=0.56$. 
   
Given that no actual asteroids associated to either of these Trojans have been detected to-date, we can reject with 95\% confidence the existence of families with $\Delta H_{1}$ less than or equal to the value for the Eureka family (2.0). This result extends, at a somewhat reduced confidence of 90\%, to cases where one of the two families contains fainter asteroids, with $\Delta H_{1}$ up to $3.0$. However, we cannot dismiss the existence of families with still fainter members. For instance, as much as 35\% of the Monte Carlo trials for families of 101429 and 121514 with $\Delta H_{1}=2.5$ yielded no detections.

The probabilities calculated for the same $\Delta H_{1}$ value for the two cases with magnitude distribution slope of $0.45$ and $0.56$ are not quite the same, with those for the steeper slope being slightly but systematically higher. It is not immediately clear why this is so, since the generation of the largest fragment involves only $\Delta H_{1}$. However, choosing a steeper slope translates into more asteroids down to a certain absolute magnitude being generated by the code. In those Monte Carlo trials where only one family member is detected, this is not always the brightest asteroid in the family but the second or third brightest. Therefore, runs with a steeper assumed slope will score slightly higher than runs with a shallower slope. 
 
Having quantified the statistical significance of our observables, we are now ready to constrain the YORPlet production rate of Martian Trojans based on the loss efficiency information obtained in Section~\ref{sec:escape}.

\section{\label{sec:constraints}Constraints from a steady state model}
\subsection{Common asteroid production rate}
We start by assuming that the Poisson rate $\lambda$ of offspring production for the three asteroids is the same and treat this as the parameter we wish to constrain. In this three-trial maximum likelihood problem, the data consists of the observed number of offspring: eight (Eureka), zero (101429) \& zero (121514). The likelihood function in this case is
\begin{equation}
\Lambda_{3}(\lambda)= \exp \left( -\lambda \sum_{i} \tau_{i} \right) \left(\lambda \tau_{3}\right)^{k}
\label{eqn:lf_three}
\end{equation}
where $i=1,\dots,3$, $\lambda$ is the - assumed common - rate of production and $\tau_{3}$, $k$ are the 
 Poisson loss rate and present size of the Eureka family respectively. The maximum likelihood estimate for $\lambda$ is
 \begin{equation}
\hat{\lambda}_{ML}=  k  / \sum_{i} \tau_{i}\mbox{, }
\label{eqn:mle_three}
\end{equation}
in other words the observed family size divided by the sum of the lifetimes. In the case at hand, $\tau_{1}$$=$0.55 Gyr while $\tau_{2}$$\simeq$$\tau_{3}$$\simeq$2 Gyr (Section~\ref{sec:escape}) where the subscript ``1'' refers to 101429, ``2'' to 121514 and ``3'' to Eureka. If we assume that each fission event creates one family member, then $k=8$ and $\hat{\lambda}_{ML}$$=$$1.7$ $\mbox{Gyr}^{-1}$. The probability of observing the sample given $\lambda=\hat{\lambda}_{ML}$ -- the statistical $p$-value -- is $1.8 \times 10^{-4}$ and $\lesssim 10^{-3}$ for $k\geq6$. On the other hand, if all of the family members were created in a single fission event i.e.~$k=1$, we obtain $\hat{\lambda}_{ML}$$=$ 0.21 $\mbox{Gyr}^{-1}$ but with a high $p$-value, $\sim$0.2. In fact, as we explore further the dependence of the result on the different assumptions on the loss timescale under $k=1$, we invariably arrive at similar values for $\lambda$ and for the associated $p$-value. This is intuitively expected since it is difficult to compare Poisson production rates based on very low event statistics. Therefore, our conclusions inevitably depend on the efficiency of producing single vs multiple offspring asteroids during a fission event. This is a poorly understood process and all we can say is that the ubiquity of asteroid pairs in the Main Belt suggests that a good fraction of fission events produce a size distribution dominated by a single, large offspring. For the remainder of the paper we focus on the $k=8$ case, returning briefly to $k=1$ near the end of  this Section in our subsequent exploration of two-trial, two-parameter tests.

Under the one-fission, one-offspring hypothesis, we now change one of our base assumptions and adopt \citet{Cuk.et.al2015} so that the orbital distribution of Eureka family members is determined by the dominance of seasonal over diurnal Yarkovsky. From Section~\ref{sec:escape}, this implies $\tau_{1}$$=$0.15 Gyr, $\tau_{2}$$=$1.6 Gyr \& $\tau_{3}$$=$ 2.9 Gyr. While we obtain the same value for the production rate, the $p$-value increases by an order of magnitude, to $3.2 \times 10^{-3}$.

We now consider the pairs 101429-Eureka and 121514-Eureka as separate two-trial problems and modify Eqs~$\ref{eqn:lf_three}$ and $\ref{eqn:mle_three}$ so that the index $i$ only runs up to 2 while the subscript ``2'' now refers to Eureka for both pairs. The $p$-value for a common offspring production rate in the former pair is 0.021 while that for the latter pair is still small, $4.9\times 10^{-4}$. Therefore, while we can generally reject the hypothesis that 121514 \& Eureka produce offspring asteroids with a common rate, we cannot do the same for the pair 101429-Eureka. Under the assumption that negative Yarkovsky is dominant we obtain 0.093 for 101429-Eureka and $4\times 10^{-3}$ for 121514-Eureka, in other words the hypothesis of common production rate for the asteroids in either pair becomes more difficult to reject.

\subsection{Independent production rates}
In the next step of our analysis, we allow the rates $\lambda_{i}$ for each pair of asteroids to vary independently, in which case the likelihood function becomes
\begin{equation}
\Lambda_{2}(\lambda_{1},\lambda_{2})= e^{-\lambda_{1} \tau_{1}} e^{-\lambda_{2} \tau_{2}} \left(\lambda_{2} \tau_{2}\right)^{k}
\label{eqn:lf_two}
\end{equation}
and is maximised for $\hat{\lambda}_{1}=0$ and $\hat{\lambda}_{2}=k \tau^{-1}_{2} $ with a $p$-value of  $\sim$$0.13$. 

\begin{figure}
\vspace{-3cm}
\centering
\includegraphics[width=85mm,angle=0]{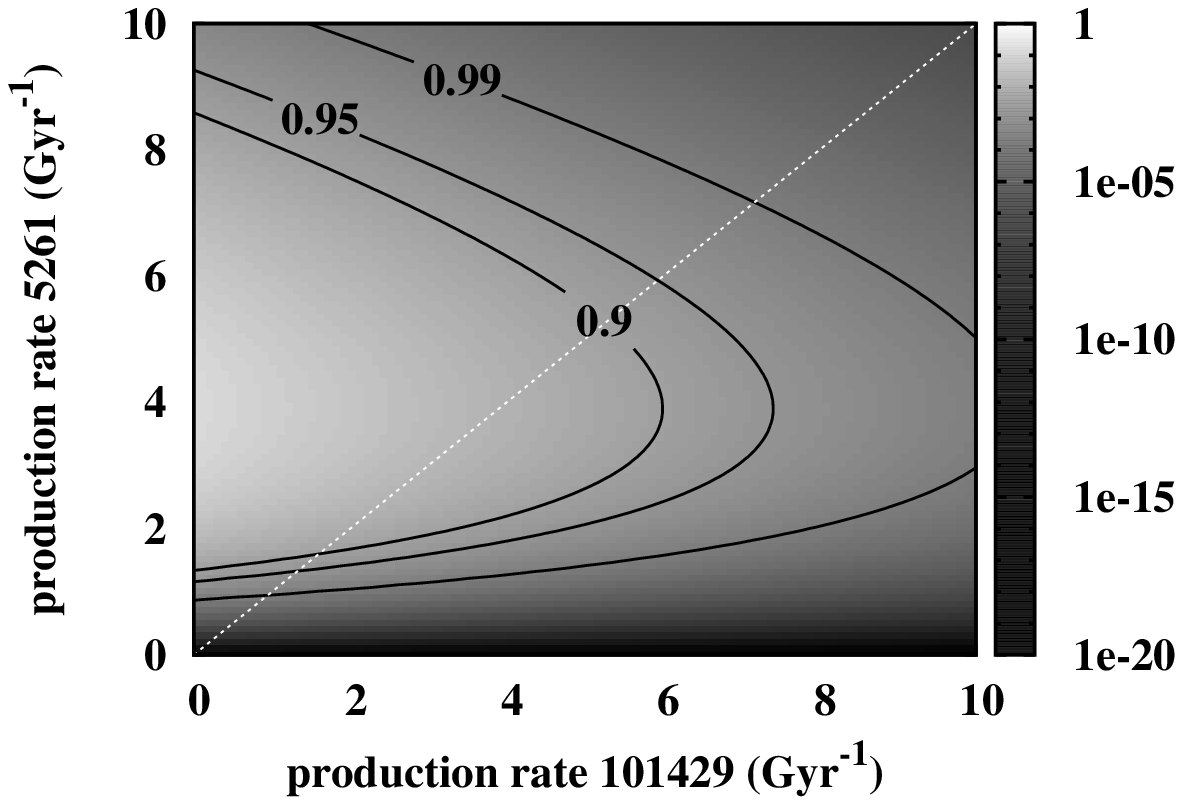}\includegraphics[width=85mm,angle=0]{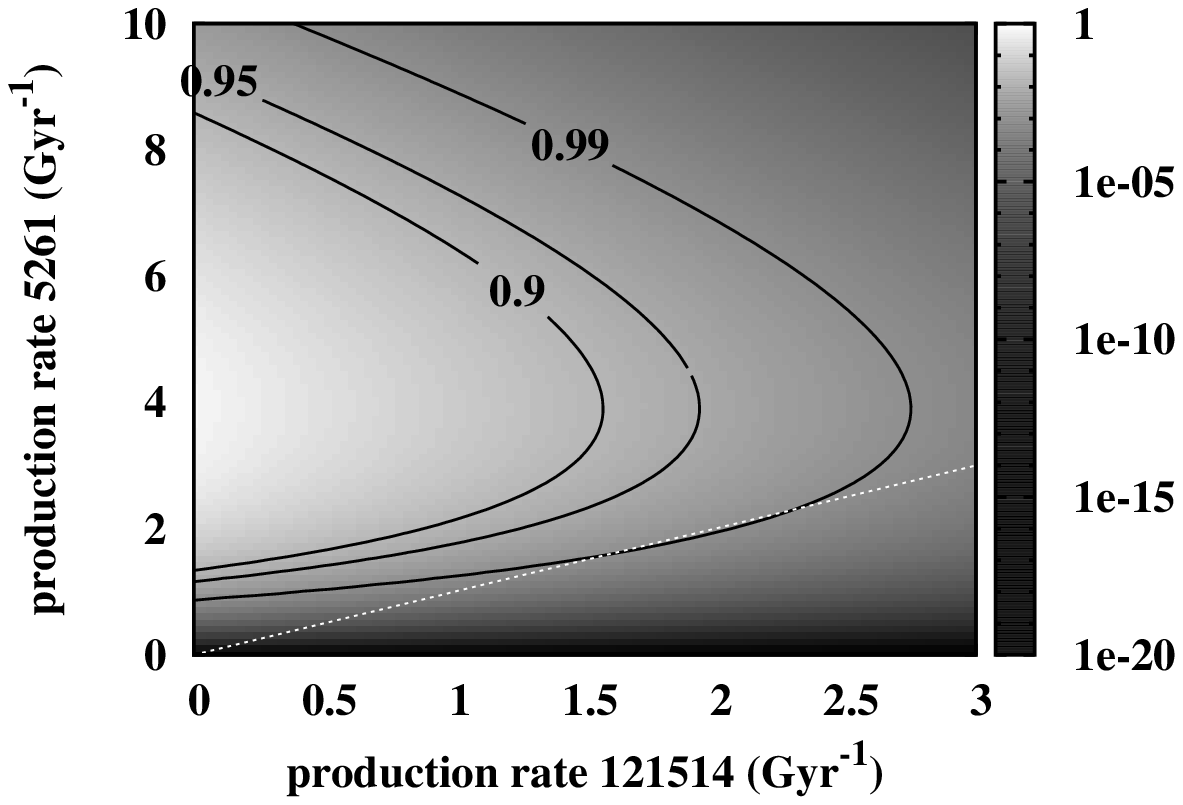}
\includegraphics[width=85mm,angle=0]{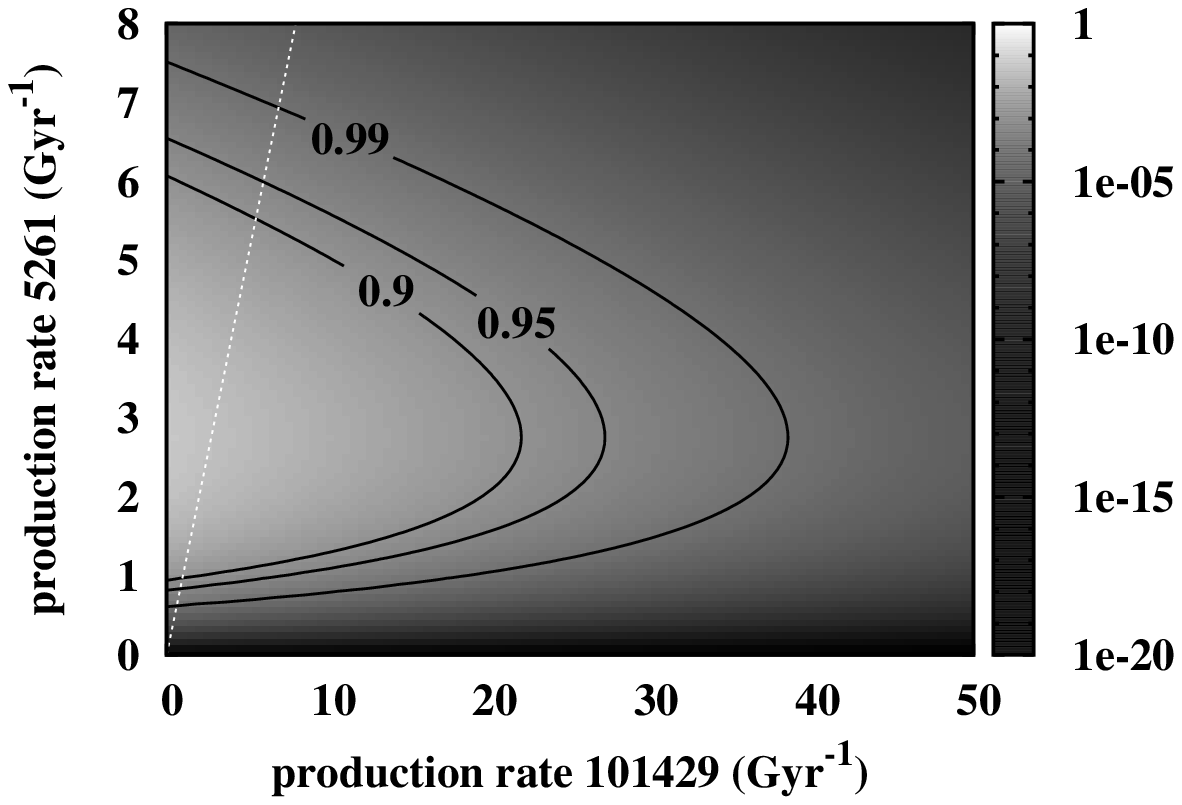}\includegraphics[width=85mm,angle=0]{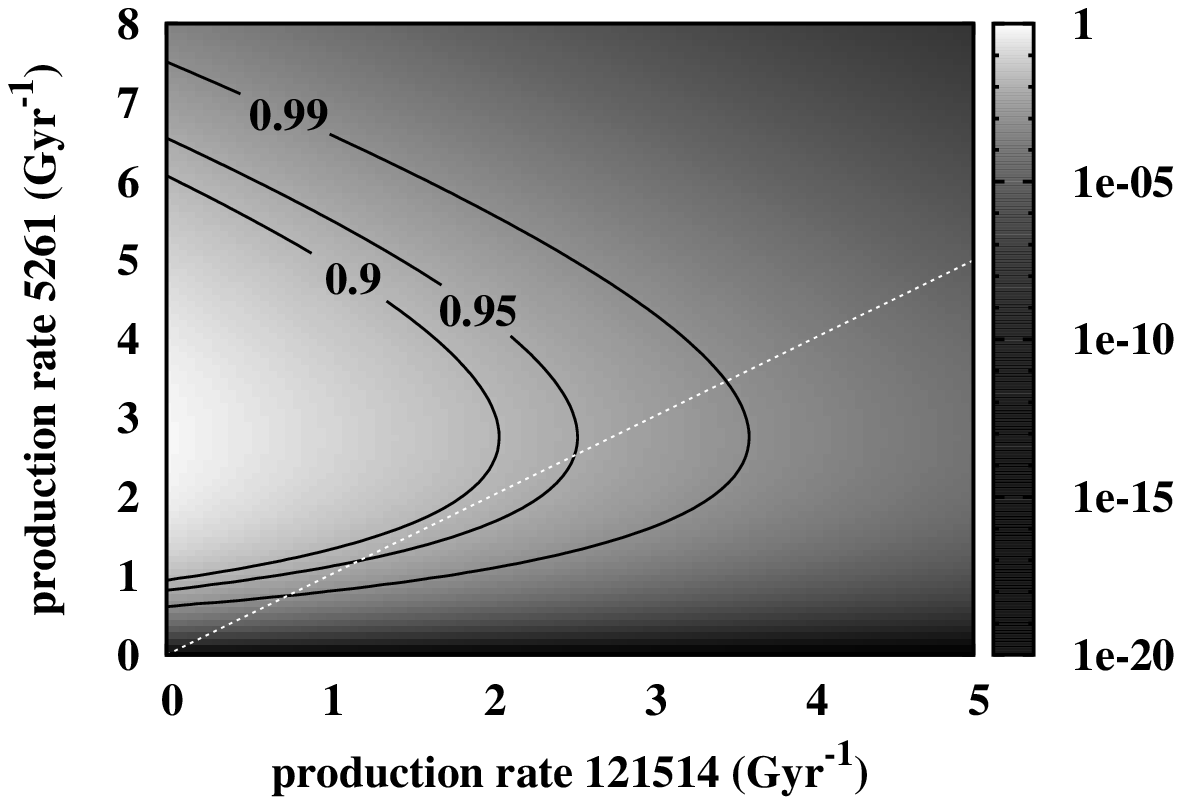}
\includegraphics[width=85mm,angle=0]{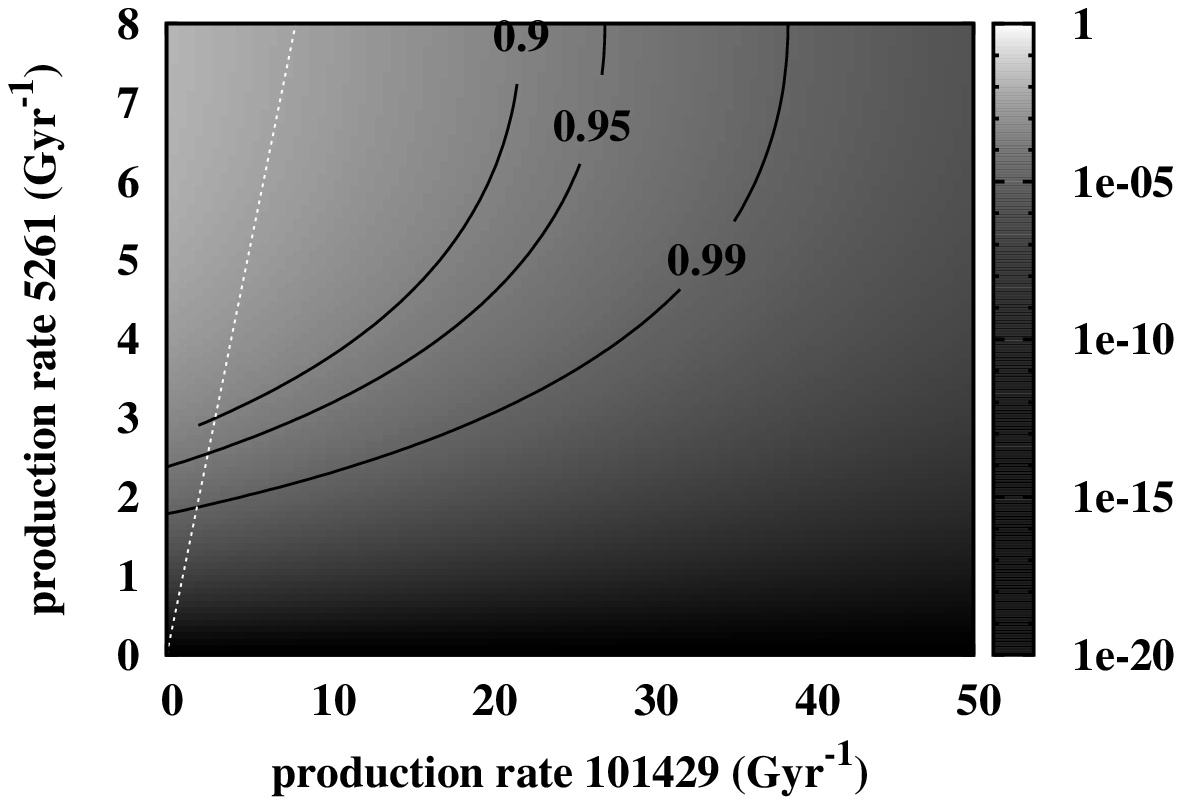}\includegraphics[width=85mm,angle=0]{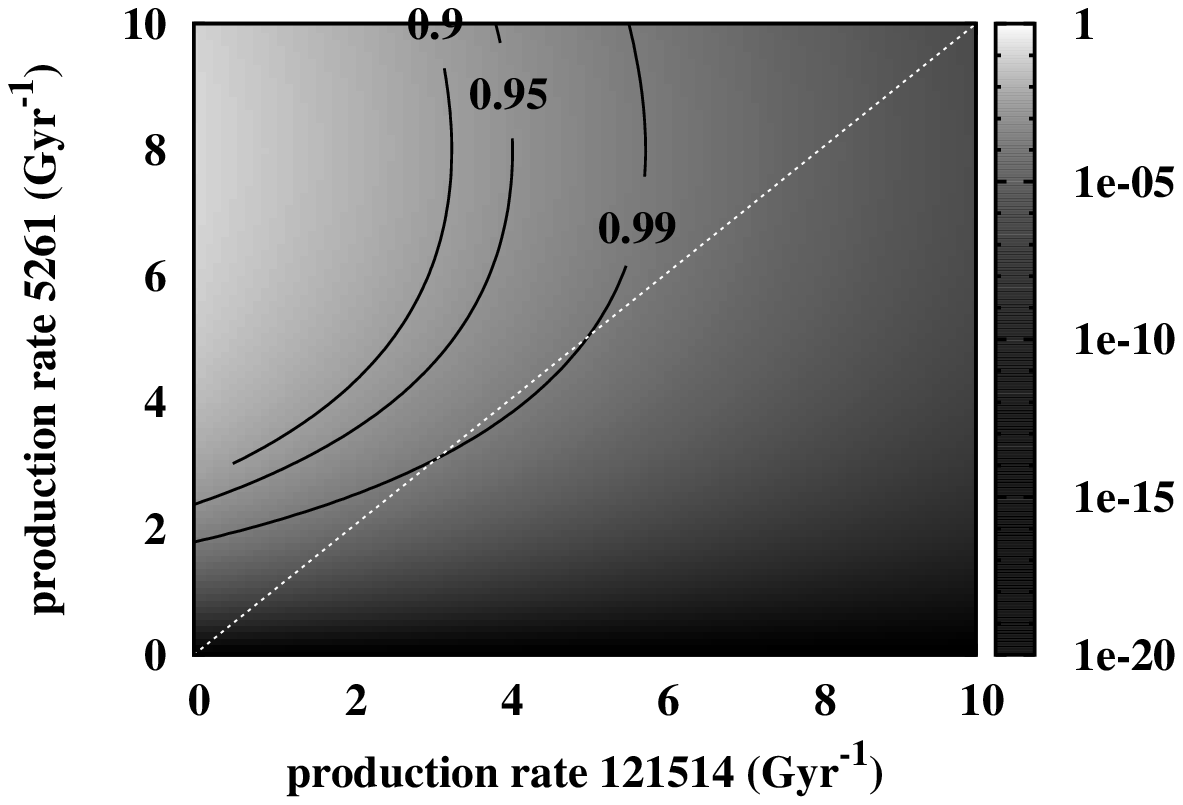}
\caption[Joint probability density of offspring production rates for 101429 (left column) and 121514 (right column) vs Eureka under different assumptions on their dynamical lifetime. Top: Asteroids are lost through Yarkovsky-driven orbital evolution independently of the sign of the acceleration. Middle: Same as top row but asteroid lifetime is determined by evolution under a negative Yarkovsky acceleration as per \citet{Cuk.et.al2015}. Bottom: As middle row but for a family lifetime determined by collisions. The white dotted line represents the locus of equal production rates for the two asteroids. The black curves demarcate confidence regions at different levels of significance.]{Joint  probability density of offspring production rates for 101429 (left column) and 121514 (right column) vs Eureka under different assumptions on their dynamical lifetime. Top: Asteroids are lost through Yarkovsky orbital evolution independently of the sign of the acceleration. Middle: As top row but asteroid lifetime is determined by evolution under a negative Yarkovsky acceleration as per \citet{Cuk.et.al2015}. Bottom: As middle row but for a family lifetime determined by collisions. The white dotted line represents the locus of equal production rates for the two asteroids. The black curves demarcate confidence regions at different levels of significance.}
\label{fig:joint_pdf_k8}
\end{figure}

For this likelihood, the probability density function (pdf) of the parameters is $\Lambda_{2}(\lambda_{1}, \lambda_{2}) \tau_{1} \tau_{2}\left(k!\right)^{-1}$ and we can use this to obtain additional information about the $\lambda_{i}$. The top panels of Figure~\ref{fig:joint_pdf_k8} show the respective pdfs for the two pairs where the adopted estimates of the escape timescale $\tau$ are those found in Section~\ref{sec:escape} under direction-independent Yarkovsky action. The black curves correspond to confidence regions at different levels of statistical significance: $>$99\%, $>$95\% \& $>$90\%. In the case of 101429, we find that the constraint imposed on the production rate $\lambda_{101429}$ is  weak: we cannot reject solutions where 101429 is producing new Trojans at a lower rate than Eureka i.e.~left of the white dotted line, even at a moderate 90\% significance level. For 121514, however, we are able to justify such a statement with 99\% significance. If we assume instead that \citet{Cuk.et.al2015} are correct in that only negative Yarkovsky is effective in changing the orbits, we obtain the result shown in the middle panels. In general, our statistical constraints weaken under this assumption. For 121514 in particular, we are still able to claim that it produces new Trojans at a lower rate than Eureka but with 90\%, instead of 99\%, confidence.

In Section~\ref{sec:escape} we found an estimated lifetime for Eureka family asteroids that is 2.5-4 times longer than the age of the Eureka family reported from the \citet{Cuk.et.al2015} simulations. This is not fatal given the factor-of-few uncertainty associated with that estimate. However, in \citet{Christou.et.al2017} and in Section \ref{sec:time} of the present paper we saw that the smallest known Trojans may experience disruptive collisions  every $\sim$Gyr and less energetic, rotational-state-resetting impacts more frequently.  Eureka family members may therefore be lost faster than assumed in our analysis. Alternatively, their orbits may disperse more slowly over time as their rotational properties, which determine the magnitude of Yarkovsky \citep{Farinella.et.al1998}, are themselves affected by the YORP cycle \citep{VokrouhlickyCapek2002,CapekVokrouhlicky2004} and by non-disruptive collisions, which result in frequent re-orientations of the spin axis and generally increase the fraction of time spent in low angular momentum states \citep{Marzari.et.al2011}. While the point regarding the rate of orbit dispersion was also made for the Eureka family by \citet{Cuk.et.al2015}, that work assumed that collisions are irrelevant in this respect whereas we saw here that this is not necessarily the case for the sub-km family Trojans. It is pertinent to ask how our conclusions will change under these circumstances, for instance if collisions with a steep-sloped impactor population remove the smaller objects over $\sim$1 Gyr. In that case, we find (bottom panels of Fig.~\ref{fig:joint_pdf_k8}) that imposing a shorter lifetime for the Trojans while maintaining our assumption of negative Yarkovsky drift as per \citeauthor{Cuk.et.al2015} strengthens the conclusion that $\lambda_{121514}<\lambda_{Eureka}$ back to 99\% significance. Similarly, the somewhat longer dynamical lifetimes (Section~\ref{sec:escape}) that result from lowering the albedo, thereby increasing the minimum observable size, of members of a 121514 family, have the effect of shifting the contours in the top two right panels of Fig.~\ref{fig:joint_pdf_k8} slightly to the left, further strengthening the statistical significance of low offspring production rate for that asteroid.   

\begin{figure}
\vspace{-3cm}
\centering
\includegraphics[width=85mm,angle=0]{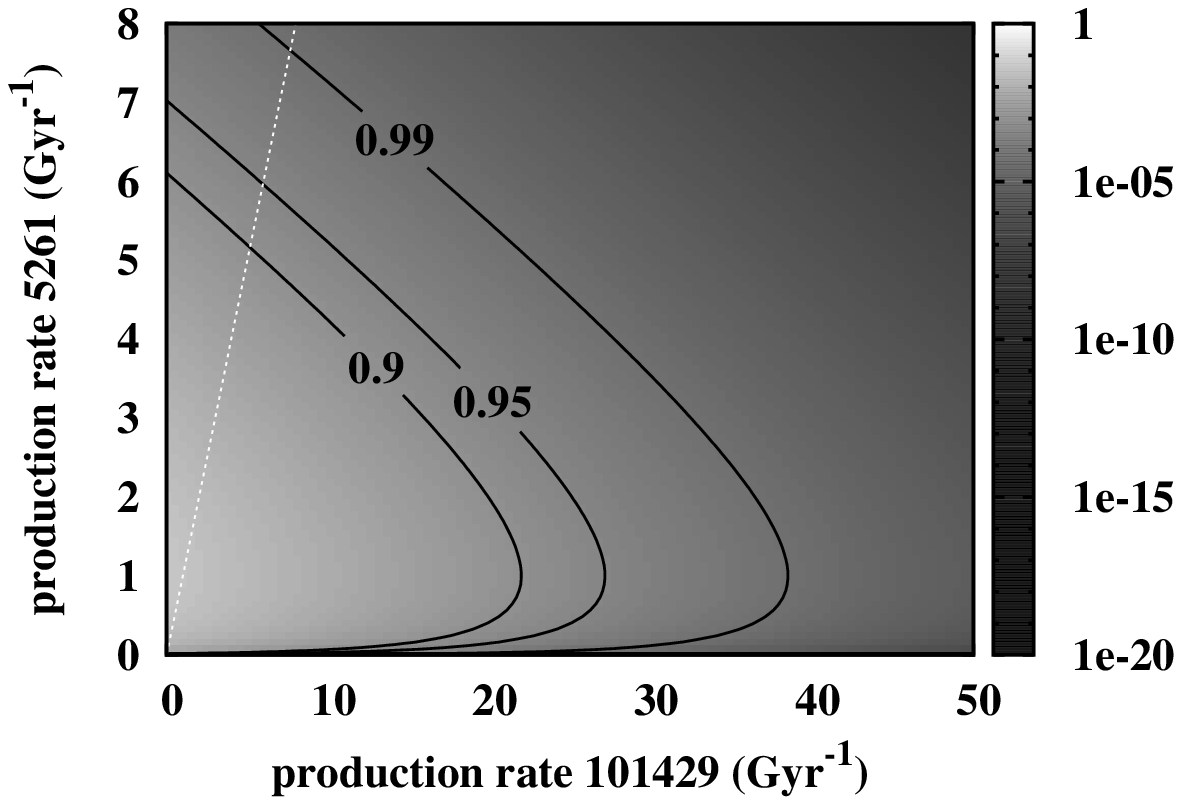}\includegraphics[width=85mm,angle=0]{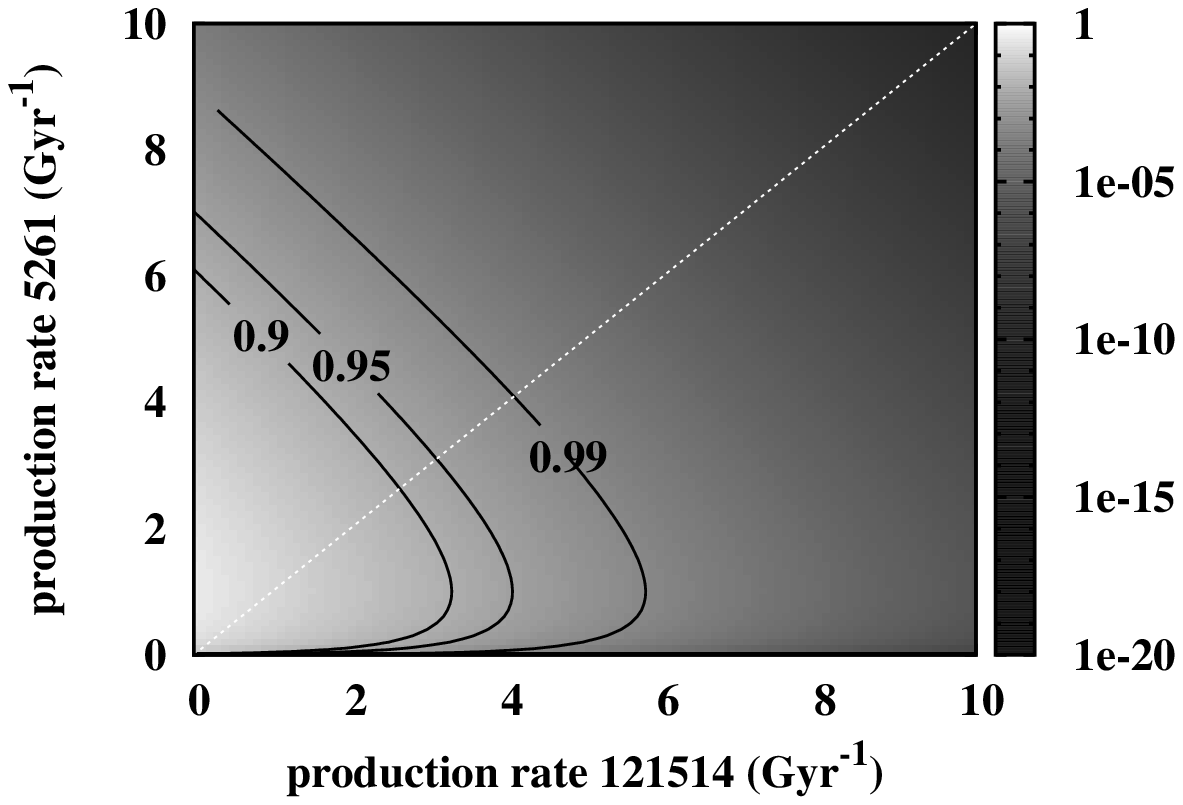}
\caption[As in bottom panels of Fig.~\ref{fig:joint_pdf_k8} but now with all Eureka family members having been created in a single fission event.]{As for bottom panels of Fig.~\ref{fig:joint_pdf_k8} but now with all Eureka family members having been created in a single fission event.}
\label{fig:joint_pdf_k1}
\end{figure}

Finally, we have carried out the same exercise for the limiting case $k=1$ where all Eureka family members are produced in a single fission event. We show in Fig.~\ref{fig:joint_pdf_k1} the result for our last set of assumptions on the loss timescale, namely that of collisional, rather than dynamical, elimination of small family members; the other two cases yield a similar result. As expected intuitively and from the earlier analysis, we are unable to obtain meaningful constraints on the rate of production from one asteroid vs the other.

In conclusion, while the unique status of Eureka among Martian Trojans can be partly explained in terms of the local dynamics of its orbital neighbourhood relative to the other large Trojans, at least one of the other large Trojans - namely 121514 - may not be an efficient producer of YORPlets. We discuss the implications and future prospects for further investigation in the next Section.

\section{\label{sec:end}Conclusions and Discussion}
The main conclusions of this paper are as follows:

\begin{itemize}
\item The observed paucity of Mars Trojan asteroids associated with 101429 at $\mbox{L}_{5}$ and with 121514 at $\mbox{L}_{4}$ is most likely not a result of observational incompleteness and indicates that the Eureka family is, in fact, the only genetic group of asteroids in the Martian Trojan clouds. This result is somewhat dependent on the absolute magnitude $\Delta H_{1}$ of the second-largest members of the putative families, if they exist, and, to a lesser degree, on their magnitude distribution slope $b$. Specifically, with $b=0.45$ we are able to reject the existence of family asteroids within the domain of either ($\Delta H^{101429}_{1}\leq 2.0$, $\Delta H^{121514}_{1}\leq2.5$) or ($\Delta H^{101429}_{1}\leq 2.5$, $\Delta H^{121514}_{1}\leq 2.0$) with 90\% confidence. These constraints persist for the slightly higher $b=0.56$ with the confidence level increasing to 95\% in this case. Therefore, if these families exist, they must have higher $\Delta H_{1}$ than the Eureka family ($\Delta H_{1}=2.0$). This is still within the range observed for small clusters in the Main Belt, thought to have arisen from YORP-induced fission \citep{Pravec.et.al2018}.
\item If the observed population of Martian Trojans is in a steady-state sustained by YORP-driven asteroid production, one object at a time, vs loss - either through Yarkovsky-driven escape from the Trojan clouds or through collisional destruction - this implies a different set of circumstances for the three parent asteroids: Eureka, 101429 and 121514. We find that the absence of asteroids associated with 101429 can be explained in terms of a high rate of loss, probably due to a nearby secular resonance at $I=30^{\circ}$ acting as an escape hatch, but that 121514 has not been producing new YORPlets at a rate as high as Eureka, at least for the last few $10^{8}$ yr and up to a Gyr.
\end{itemize}

The picture proposed here for the Martian Trojans may be too simple for a number of reasons. First of all, our assumption that Eureka is the only object producing new observable family members may not, in fact, hold; asteroids 311999 ($H=18.1$) and 385250 ($H=18.9$) are large enough to have themselves been parents of one or more of the other family members by YORP-induced fission and subsequent escape \citep{Scheeres2009,Pravec.et.al2010,Christou2013}. In addition, we saw in Section~\ref{sec:escape} that the stability of 101429 as a Martian Trojan is dependent on the strength, but also the direction, of the Yarkovsky force applied to it. The most stable clones in our simulations are those that evolve under positive Yarkovsky acceleration rather than those with the smallest acceleration magnitude. Therefore, if this object is a long-term resident, the net Yarkovsky acceleration on its orbit was either zero or within a relatively narrow range of positive values. Alternatively, it may have been captured from heliocentric orbit relatively recently ($\ll \tau$), perhaps with the Yarkovsky effect acting to stabilise the orbit post-capture instead of acting to destabilise it. In that case, the absence of a family would be a consequence of this asteroid's recent arrival to the Martian Trojan clouds rather than the efficient escape of offspring through the secular resonance\footnote{This follows from setting $m=0$ in Eq~\ref{eq:ex} of the Appendix and noting that, for $t\ll\tau$, $E[x] \ll \lambda \tau$}. Long-term evolution under positive Yarkovsky for this Martian Trojan would be at odds with the conclusion of \citet{Cuk.et.al2015} that the seasonal component dominates over the diurnal one over Gyr timescales for the Eureka family. This is not problematic for the recent capture hypothesis, since Yarkovsky would have acted over a timescale much shorter than a Gyr.

The apparent inability of 121514 to efficiently produce YORPlets may be related to its low spin rate. In the Main Belt, YORP may drive a significant fraction of km-sized or smaller asteroids to spin states characterised by very long rotation periods \citep{Rubincam2000,VokrouhlickyCapek2002,CapekVokrouhlicky2004} and/or chaotic wandering of the rotation axis, the so-called tumbling state \citep{Vokrouhlicky.et.al2007} which may persist until internal energy dissipation or an external factor such as a collision \citep{Marzari.et.al2011} re-instates principal axis rotation. \citet{Borisov.et.al2018} noted that the characteristic lifetime for a body of the size and rotation rate of 121514 to re-enter into principal axis rotation from a tumbling state through tidal dissipation is a significant fraction of the age of the solar system. From Section~\ref{sec:time}, it does not appear likely that a collision energetic enough to reset the asteroid's rotational angular momentum would have taken place in the last Gyr or so of the asteroid's existence at 1.5 au from the Sun. Therefore, 121514 may be an example of a ``barren'' asteroid, trapped for {\ae}ons in a low-angular-momentum state from which it has been unable to escape.  
 
Apart from advances in theoretical modelling to better understand YORP fission outcomes \cite[eg ][]{CottoFigueroa.et.al2015} and evolution in the 1:1 mean motion resonance under the Yarkovsky effect \citep{Hou.et.al2016,WangHou2017}, new observations will be crucial to making further progress. Targeted investigations of the brightest Trojans are highly desirable to improve our knowledge of these objects' physical properties such as size, shape and rotational state. In the short-term, it would be useful to confirm the tumbling state of 121514 and to check whether 101429 rotates in a prograde or retrograde sense since this determines the sign of the Yarkovsky acceleration acting on it. In the longer term, facilities such as PanSTARRS \citep{Jedicke.et.al2007,Wainscoat.et.al2015}, GAIA and, especially, the Large Synoptic Survey Telescope \citep{Jones.et.al2015} will improve the current level of population completeness of the Trojans (shown here to be $H_{lim} \simeq 20$) by several magnitudes and should discover a few hundred additional members of the Eureka family (Christou, A.~A., {\it IAU 2018 Coll.~Proc.}, in press). This will either confirm the lack of families of 101429 and 121514 or find their brightest members, if they exist. In addition, because the path in the space of libration amplitude vs inclination is deterministic \citep{Cuk.et.al2015}, an approach similar to those used in \citet{Milani.et.al2017} and \citet{BrozMorbidelli2019} to place bounds on ages of asteroid families in resonances will further constrain the Eureka family age and help remove one of the principal unknowns in the effort to understand the Martian Trojans.
 \ack
The authors wish to thank Miroslav Bro\v{z} and an anonymous reviewer whose suggestions considerably improved the manuscript. Work by AAC, GB, AC \& AD reported in this paper was supported via grants (ST/M000834/1 and ST/R000573/1) from the UK Science and Technology Facilities Council (STFC). We acknowledge the SFI/HEA Irish Centre for High-End Computing (ICHEC), the Dublin Institute for Advanced Studies (DIAS) as well as the University of Florida (UF) Department of Astronomy for the provision of computational facilities and support. This publication makes use of data products from the Wide-field Infrared Survey Explorer, which is a joint project of the University of California, Los Angeles, and the Jet Propulsion Laboratory/California Institute of Technology, and NEOWISE, which is a project of the Jet Propulsion Laboratory/California Institute of Technology. AAC gratefully acknowledges support (Short Visit Grant \#6231) from the Gaia Research in European Astronomy Training collaboration within the framework of the European Science Foundation. Astronomical research at the Armagh Observatory and Planetarium is grant-aided by the Northern Ireland Department for Communities (DfC).\label{lastpage}

\bibliography{Christou_etal_icarus_2019}
\bibliographystyle{plainnat}
\clearpage
\protect
\listoffigures

\renewcommand{\baselinestretch}{1.0}
\protect

\newpage
\begin{table}[htb]
\centering
\caption[Physical properties of Martian Trojans.]{Physical properties of Martian Trojans.}
\begin{tabular}{lcccc}
\noalign{\smallskip}
\hline \hline
\noalign{\smallskip} 
                   &  Absolute  & Diameter &  Period  & Visible \\
Designation        & Magnitude & (km) & (hr) & albedo \\\hline \noalign{\smallskip}
(5261) Eureka   &     16.1  &    1.878$\pm$0.231    &     2.69  &  0.182$\pm$0.054 \\    
(101429) 1998 $\mbox{VF}_{31}$  & 17.2   &  $\mbox{0.946}^{\ast}$  & $\mbox{4.67-9.34}^{\dagger}$ &  $\mbox{0.26}^{\ast\ast}$ \\
(121514) 1999 $\mbox{UJ}_{7}$    &    17.1   &   2.286$\pm$0.495 & $\mbox{46.5}^{\ddagger}$  &   0.053$\pm$0.034  \\ \noalign{\smallskip} \noalign{\smallskip} 
\hline \hline  \noalign{\smallskip} 
\multicolumn{5}{l}{\parbox{119mm}{
Data retrieved on 01 Oct 2018 from the JPL Small-Body Database Browser (https://ssd.jpl.nasa.gov).${}^{\ast}$Calculated from $H$.${}^{\ast\ast}$Average for S-type  asteroids from \citet{DeMeoCarry2013}.${}^{\dagger}$Based on re-analysis of lightcurve from \citet{Borisov.et.al2016}.${}^{\ddagger}$Rotation period from \citet{Borisov.et.al2018}.}}
\end{tabular}
\label{tab:troj}
\end{table} 

\newpage
\begin{table}[htb]
\centering
\caption[Parameters of the numerical simulations of Trojan orbital evolution.]{Parameters of the numerical simulations of Trojan orbital evolution.}
\begin{tabular}{lccc}
\noalign{\smallskip}
\hline \hline
\noalign{\smallskip} 
                   &  Duration  & Number&  Range of $\dot{a}$${}^{\dagger}$  \\
Designation        & (Gyr) & of clones & ($\times10^{-3}$ au $\mbox{Myr}^{-1}$) \\\hline \noalign{\smallskip}
(5261) Eureka   &     3.6  &    50    &   $-$2.0 $\rightarrow$ + 2.0 \\    
(101429) 1998 $\mbox{VF}_{31}$  & 2.0   &  50 &    $-$2.0 $\rightarrow$  + 2.0 \\
(121514) 1999 $\mbox{UJ}_{7}$    &    3.6   & 30   &    $-$2.0  $\rightarrow$  + 2.0 \\
\noalign{\smallskip} \hline \hline  \noalign{\smallskip} 
\multicolumn{4}{l}{\parbox{117mm}{
${}^{\dagger}$The boundary values correspond to the semimajor axis drift rate for an asteroid with radius 150 m, visible albedo of $0.2$ and a rotation period of 4 hr, according to the diurnal Yarkovsky model of \citet{Farinella.et.al1998}.}}
\end{tabular} 
\label{tab:sims}
\end{table}

\newpage
\begin{table}[htb]
\centering
\caption[Parameters of the Monte Carlo simulations.]{Parameters of the Monte Carlo simulations.}
\begin{tabular}{ccc}
\noalign{\smallskip}
\hline \hline
\noalign{\smallskip} 
& Calibration & Survey \\\hline \noalign{\smallskip}
 $T_{0}$ (yyyy-mm-dd) & 1998-06-01 &  " \\
 $T_{1}$ (yyyy-mm-dd) & 2018-06-01 &  " \\
Number of Monte Carlo trials ($N_{t}$) &  1000  & 1000\\
Number of test observation epochs ($N_{e}$) &  1000  & 100\\
Number of test orbits ($N_{o}$$\times$$N_{o}$) & 30$\times$30 & 1\\    
$V_{50}$ & 21.0 & " \\
w  &  0.2  & " \\
 $E_{lim}$  &   $70^{\circ}$ &   \\
\# detections to discover ($n_{d}$) &  $\geq$41 & $\geq$4   \\
\noalign{\smallskip} \hline \hline  \noalign{\smallskip} 
\end{tabular} 
\label{tab:sims2}
\end{table} 

\newpage
\begin{table}[htb]
\centering
\caption[Cumulative distribution of the bivariate probability that hypothetical families of 101429 and 121514 exist given that no members of either family have so far been detected. The slope of the assumed $H$ distribution for a family is $b$$=$0.45, the same as our estimate for the Eureka family.]{Cumulative distribution of the bivariate probability that hypothetical families of 101429 and 121514 exist given that no members of either family have so far been detected. The slope of the assumed $H$ distribution for a family is $b$$=$0.45, the same as our estimate for the Eureka family.}
\begin{tabular}{cccccc}
\noalign{\smallskip}
\hline \hline
\noalign{\smallskip} 
&\multicolumn{5}{c}{121514} \\\hline \noalign{\smallskip}
& $\Delta H_{1}\leq$      & 1.5               & 2.0             & 2.5               &        3.0   \\
& 1.5 &   {\bf 0.0008}      &   {\bf 0.0056}     &   {\bf 0.0207}     &    {\bf 0.0485}  \\
101429 & 2.0 &    {\bf0.0032}     &   {\bf 0.0229}     &  {\it 0.0850}      &     0.1989     \\
& 2.5 &     {\bf 0.0088}    &  {\it 0.0628}      &   0.2336     &   0.5463     \\
& 3.0 &     {\bf 0.0162}    &     0.1149644    &   0.4275     &    1.0000   \\    
\noalign{\smallskip} \hline \hline  \noalign{\smallskip} 
\multicolumn{6}{l}{\parbox{117mm}{Values in bold and italics correspond to $>$95\% and $>$90\% significance respectively.}}
\end{tabular} 
\label{tab:sign_0p44}
\end{table} 

\newpage
\begin{table}[htb]
\centering
\caption[Same as Table~\ref{tab:sign_0p44} but for $b$$=$0.56 obtained from a fit to the \citet{Pravec.et.al2018} data on Main Belt asteroid clusters.]{Same as Table~\ref{tab:sign_0p44} but for $b$$=$0.56 obtained from a fit to the \citet{Pravec.et.al2018} data on Main Belt asteroid clusters.}
\begin{tabular}{cccccc}
\noalign{\smallskip}
\hline \hline
\noalign{\smallskip} 
 & \multicolumn{5}{c}{121514}  \\\hline \noalign{\smallskip}
& $\Delta H_{1}\leq$    & 1.5               & 2.0              & 2.5              & 3.0         \\
& 1.5 &   {\bf 0.0000}      &   {\bf 0.0006}     &   {\bf 0.0041}     &    {\bf  0.0120}  \\
101429 & 2.0 &   {\bf  0.0002}     &    {\bf 0.0059}    &  {\bf 0.0427}     &     0.1239  \\
& 2.5 &    {\bf  0.0007}    &  {\bf 0.0230}     &   0.1669     &   0.4835   \\
& 3.0 &     {\bf 0.0016}    &    {\bf 0.0477}   &   0.3451     &    1.0000  \\     
\noalign{\smallskip} \hline \hline  \noalign{\smallskip} 
\end{tabular} 
\label{tab:sign_0p56}
\end{table}

\clearpage

\clearpage

\clearpage

\clearpage

\clearpage

\clearpage

\clearpage

\clearpage

\clearpage

\clearpage
\appendix
\section{\label{sec:app1}Production and loss of Trojans as a random process: Problem formulation and analytical solution}

The problem may be modelled as a birth-death stochastic process\footnote{Equivalent to an M/M/$\infty$ zero-waiting-time queueing problem \cite[][Chapter 11]{AdanResing2002}. If the lifetime of each offspring is deterministic and fixed at the same value $\tau$ (an M/D/$\infty$ queue), the distribution of $x$ is the same as for the steady state of the stochastic case.} that supports $x(t)$ offspring asteroids at time $t$ with some probability $p_{x}(t)$. We assume that the observed offspring asteroids are produced in a Poisson process where the probability of birth in a ``short'' interval of time $h$ is $ \lambda h + o(h)$ where $o(h)/h$ vanishes as $h \rightarrow 0$. Similarly, an offspring Trojan asteroid may be lost with a probability $ \tau ^{-1} x h$ where $\tau$ is the characteristic lifetime of the offspring.

The sequence $p_{x}=P(x(t)=x)$ defines the probability density function of $x(t)$. It satisfies the system of ODEs:
\begin{eqnarray}
p^{\prime}_{0}& = & - \lambda p_{0} + \tau^{-1} p_{1} \\
p^{\prime}_{x} & = &- \left(\lambda + x \tau^{-1} \right) p_{x} + \lambda p_{x-1} +(x+1) \tau^{-1} p_{x+1}\mbox{, }x=1,2,\cdots
\end{eqnarray}
the solution to which may be indirectly obtained by multiplying both sides by $\theta^{k}$ to form the probability-generating function (pgf) $ P(\theta, t) = \sum_{k=0}^{\infty} p_{k} \theta^{k}$ which satisfies
\begin{equation}
\frac{\partial P}{\partial t} = - \lambda ( 1 - \theta) P + \tau^{-1} ( 1 - \theta) \frac{\partial P}{ \partial \theta}
\end{equation}

the general solution to which is
\begin{equation}
P(\theta,t) = {(1 - e^{-t/\tau}(1 - \theta))}^{m}e^{-\lambda  \tau (1 - \theta) (1 - e^{-t/\tau})} \mbox{ where  $x(0)=m$.}
\end{equation}

The steady state solution may be recovered by setting $t \rightarrow \infty$ to obtain
\begin{equation}
P(\theta, \infty) = e^{-\lambda  \tau (1 - \theta)}
\label{eq:steadystate}
\end{equation}

 i.e. a Poisson process with rate $\lambda \tau$. Note that the probability of $x(t)=0$ converges asymptotically to $ e^{-\lambda \tau} $ independently of $m$.

The first two moments of the distribution of $p_{x}$ are
\begin{eqnarray}
E[x(t)]&=&\lambda \tau (1- e^{-t/\tau}) + m e^{-t/\tau} \label{eq:ex}\\
V[x(t)]&=&E[x(t)] - m e^{-2 t/\tau} \label{eq:vx}
\end{eqnarray}

For $t \rightarrow \infty$, we recover the expectation and variance of the distribution (\ref{eq:steadystate}).    

\section{\label{sec:app2}Comparison between orbit integrations of \citet{Cuk.et.al2015} and this work}
Here we compare our simulations in Section~\ref{sec:escape} with those of \citet{Cuk.et.al2015} who simulated the orbit evolution of the Eureka family under Yarkovsky. The two computational implementations of the Yarkovsky force are independent. In addition, \citeauthor{Cuk.et.al2015} used $|\dot{a}|\leq 1.4 \times 10^{-3}$ au $\mbox{Myr}^{-1}$ for their clone simulations compared to $|\dot{a}|\leq 2 \times 10^{-3}$ au $\mbox{Myr}^{-1}$ used in this work. To compare their results with ours, we utilise those authors' conclusion that the evolution of the inclination and the libration amplitude of Eureka family members is regular and dominated by the Yarkovsky force. We choose as a proxy quantity the orbital inclination of those clones that suffered maximum orbit change over 1 Gyr. In their simulations (cf their Figs 3 \& 6) the inclination changes from $\sim$$22^{\circ}$ to $\sim$$18^{\circ}$ for clones affected by negative Yarkovsky and from $\sim$$22^{\circ}$ to $\sim$$27^{\circ}$ for those affected by positive Yarkovsky. The corresponding changes we observe in our runs are from $\sim$$22^{\circ}$ to $\sim$$16^{\circ}$ in the first instance, and from $\sim$$22^{\circ}$ to $\sim$$31^{\circ}$ in the second. We conclude that the difference in orbit evolution arising from our use of a slightly higher value of the maximum non-gravitational acceleration strength is less than a factor or two for those clones that evolved under negative Yarkovsky.
\end{document}